\documentclass[12pt,plainnat]{elsarticle}

\usepackage{amssymb}
\usepackage{amsmath}
\usepackage{url}
\usepackage{soul}
\usepackage{geometry}
 \usepackage{graphbox}
\usepackage{xcolor}

\journal{Mechanical Systems and Signal Processing}

\newcommand{\keff}{k_\mathrm{eff}}

\newcommand{\Tr}{T_{ref}}
\newcommand{\kfs}{k_{\mathrm{fs}}}
\usepackage{caption}

\begin{document}
\begin{frontmatter}

\title{Self Calibration by ON/OFF Sensitivity Switching - Feasibility Study of a Resonant Accelerometer}

\author{A. Zobova, S. Krylov}

\affiliation{organization={Faculty of Engineering, School of Mechanical Engineering, Tel-Aviv University},
            city={Ramat Aviv},
            postcode={6997801}, 
            country={Israel}
            }

\begin{abstract}

This research provides the theoretical feasibility study of a novel architecture of a microelectromechanical (MEMS) differential resonant accelerometer, with switchable and tunable electrostatic transmission between the proof mass and the vibrating sensing beams. The same beams  are used for sensing of the inertial force, while the transmission is switched ON, and for the device's  calibration, when the transmission  is OFF. Therefore, the beams' response in the OFF state is affected by the same factors (temperature, electronics, packaging) as in the ON state, with the only exception for the acceleration. This unique ability to physically disconnect the inertial force from the sensing elements opens possibilities for new schemes of the signal processing, including sensitivity tuning, zero-bias correction and on-the-fly self-calibration of the sensor.
The device includes a proof mass and two force-transmitting frames that are attached to the substrate by the suspension springs, such that there is no direct mechanical connection between the mass and the frames. Two identical sensing beams (resonators) are attached at their ends to the frames and the substrate.  When the electrostatic transmission is switched ON by  applying a control voltage between the proof mass and each frame, the axial force is transmitted from the proof mass through the frame to the resonant beam. Disturbance in the electrostatic field, due to the acceleration-dependent displacement of the proof mass, results in the shift in the beam axial force and, therefore, in its resonant frequency, assuring the device's acceleration sensing. Furthermore, the change in the control voltage tunes the transmission of the input signal, and therefore the scale factor and the dynamic range of the sensor. An electro-mechanical lumped model of the generic device is formulated and verified using full-scale multiphysics finite elements analysis. The tunability of the device and the compensation of the scale factor thermal sensitivity are demonstrated using the model.

\end{abstract}


\begin{highlights}
\item Novel architecture and operational principle of micro-electromechanical inertial sensor with tunable and ON/OFF switchable electrostatic transmission of the inertial signal to the sensing element is introduced.
\item A feasibility study of a resonant differential accelerometer is conducted by means of analytical and finite element models.
\item Temperature compensation approach is suggested.
\end{highlights}


\end{frontmatter}
\newpage

\section{Introduction} \label{sec:intro}
Microelectromechanical accelerometers are among the most widely implemented and commercially successful microdevices, due to their low power consumption, low cost, and expanded range of integration possibilities~\cite{wang2020micromachined,zhang2024review}. Among the most promising approaches, allowing performance enhancement frequency, monitoring-based sensing is intensively explored. In resonant accelerometers,  the dependence of the sensing element's natural frequency on the inertial (acceleration) force acting on the proof mass is exploited. The advantages of resonant sensors over more widely adopted statically operated devices include a wider dynamic range, higher bias stability and reduced 1/f noise~\cite{li2007ultra}. The simplest architecture of a resonant sensor of this kind is a  vibrating beam accelerometer (VBA), in which the  beams are directly or indirectly attached to the proof mass and are stretched or compressed by an inertial force~\cite{Hopkins1998TheGuidance}.
In the framework of a commonly adopted differential sensing scenario~\cite{roessig1998integrated,zhang2023thermal}, the acceleration is extracted by measuring the difference in resonant frequencies between the two (stretched and compressed) elements.
Since in these devices the force transferred to the sensing beams by the proof mass and, consequently, the scale factor (SF) are small,  resonant accelerometer designs, implementing compliant force amplification mechanisms, were reported~\cite{seshia2002vacuum,su2005resonant,shin2017environmentally}. 

One of the key aspects influencing  performance of sensors in general, and resonant accelerometers in particular, is a trade-off between their sensitivity (the SF)  and dynamic range~\cite{babatain2021acceleration}. 
In resonant accelerometers, an ability to increase sensitivity, for example by using higher force amplification, is limited by the SF nonlinearity~\cite{zhang2024review}.
One of the approaches suggested to overcome this fundamental difficulty is the use of electrostatic (ES) tuning of the sensing resonator frequency and the device's SF~\cite{comi2016sensitivity,peng2019sensitivity,Shin2019}. This approach opens a possibility to tailor the figures of merit to fit specific purposes~\cite{wang2020micromachined}. In addition, the voltage  allows control of the frequency split between the two resonant sensing elements (which are never identical due to fabrication tolerances~\cite{peng2019sensitivity}) in the cases when differential measurement is implemented. However, electrostatic forces used for  the resonant sensing elements tuning may also interact with the inertial signal and affect the SF and its nonlinearity. As a result, a need to separate the ES and inertial signals in resonant accelerometers imposes an additional design challenge. 

Another important figure of merit dictating the performance of   resonant accelerometers is  the dependence of the sensor's  SF on temperature. This dependence  emerges due to several reasons, including  the intrinsic dependence of the elastic modulus on temperature  (quantified by the temperature coefficient of elasticity, TCE),  residual stress which originated in the fabrication process or in the thermal mismatch of multimaterial packaging assemblies.  In principle, the dependence of the frequency on temperature can be eliminated by using an intrinsic common mode rejection feature of a differential measurement scenario \cite{roessig1998integrated,zhang2023thermal}.  Indeed, in the framework of the differential sensing, the frequency shifts of the two 
identical resonators, due to the changes in the temperature, are the same and the output of the sensor is not affected by the thermal effects. However, the advantages of the differential approach are partially hindered by fabrication tolerances, when two resonators are not perfectly identical, or in the case of a non uniform in spatial temperature distribution. In addition, the temperature affects not only the mechanical core of the sensor, but also the electronic circuit and its elements, such as resistors, capacitors, amplifiers and others~\cite{gonzalez2006foundations, delke2021single}.

To compensate for these dependencies and enlarge the temperature working range, the table look-up compensation approach, combined with the on-chip thermoresistors for thermal measurements, are used~\cite{shin2021temperature,wang2023temperature,Ma2024,cai2021improved}. 
However, the table look-up approach complicates the fabrication and the calibration of the device. The inevitable time delay between the thermo-resistor signal and the sensing output may deteriorate the performance \cite{Ma2024}. In addition, thermal field non-uniformity, combined with location of thermo-resistors at certain distance from the sensor, may introduce errors.  
The novel scheme combining two acceleration-sensitive and temperature-sensitive resonators on the same chip, with an improved scheme of signal processing,  was proposed in \cite{Ma2024}: it allows compensating for the temperature zero-drift without time delay; however, the increasing nonlinearity narrows the sensor's working range.

One of the central figures of merit of the resonant accelerometer, to a large extent defining the device performance, is its bias stability~\cite{Barbour2001332}.
Among a large variety of possible solutions suggested in an attempt to reduce drifts and to improve bias stability of inertial sensors, self-calibration emerged as one of the promising and intensively investigated approaches allowing to mitigate drifts. A common self-calibration approach includes applying an artificial external stimulus emulating the quantity of interest.
This is achieved, for example, by mounting the inertial sensor on a moving or rotating platform with known acceleration or angular rate \cite{puers2002rasta,aktakka2014microactuation,nadig2016run,nadig2018situ,chen2018chip,cui2019effective,benjamin2023electrostatically}, using multiple inertial measurement units (IMUs) simultaneously \cite{larey2020multiple},  or applying an additional electrostatic force emulating the inertial force \cite{ozel2016electrical,sabater2022towards}. Note that the performance of the calibrating systems still can be affected by the uncertainties of the temperature-dependent emulating
forces.
Within the self-calibrating arena, a zero signal or zero velocity updating algorithms (ZUPT) are based on the general {\it a priori} knowledge of the performed task. The algorithms recognize the time intervals where the signal is zero,  allowing updating the inner phase variables during these intervals \cite{yang2004analysis,foxlin2005pedestrian,ojeda2007non,glueck2013real,wang2018analytical,wang2019research,zhao2021pseudo,jao2022neural}.
The ZUPT was successfully implemented in ground vehicles, where the
inertial navigation is enhanced by using odometry, or in pedestrian navigation systems, where zero velocity/acceleration
events are detected using accelerometers mounted on the human body. By implementing this approach, up to  94\% decrease in the positioning error in pedestrian navigation systems was achieved~\cite{jao2022neural}.
However, these odometry-based approaches cannot be implemented on airborne or autonomous submersible platforms. 

We can conclude, therefore, that there is a need in the development of new 
architectures and operational scenarios 
to reach wide-range SF tunability  (to overcome the intrinsic trade-off between the scale factor and the dynamics range, as well as to compensate for fabrication tolerances) on the one hand, and to provide self-compensation abilities, on the other hand. Self-calibration abilities based on the ZUPT approach 
and adopted for the implementation in non-odometric systems can serve as an efficient tool for temperature compensation and zero-acceleration bias reduction.

\begin{figure}[t]
    \centering
    \includegraphics[width = 0.65\textwidth]{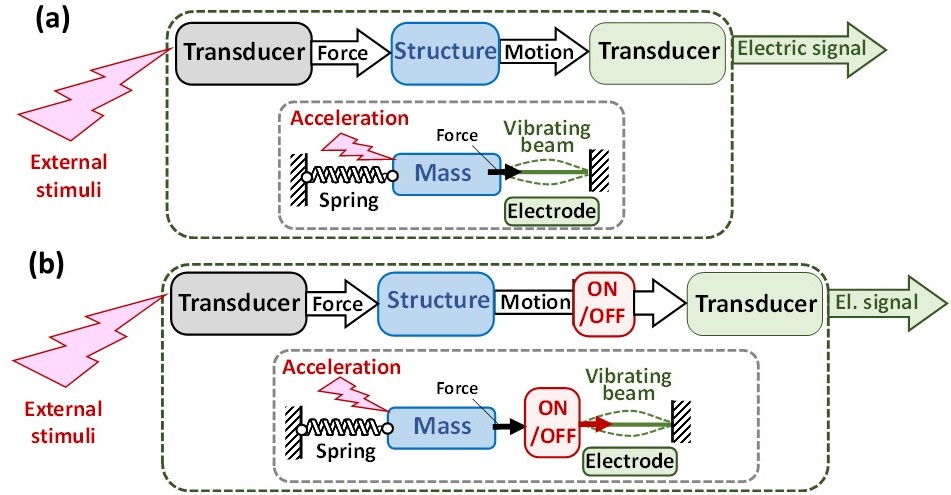}
    \caption{(a) Generic architecture of an electromechanical sensor. (b) A generic senor enhanced by an ON/OFF switchable (or/and tunable) transmission of the mechanical signal to the sensing element.  Insets represent an example of a resonant accelerometer with a vibrating beam as a sensing element.}
    \label{fig:concept}
\end{figure}

Here we introduce an approach based on the active tuning and ON/OFF switching of the inertial signal transferred to the sensing element of the sensor as a tool for the sensor SF tuning and thermal compensation. On the most basic level,  most of the electromechanical (as opposed to purely optical, chemical or electronic sensors) sensors rely on the same operational scenario, Fig.~\ref{fig:concept}(a). Namely, an environmental signal causes, directly, or as a result of a corresponding transduction, mechanical displacement or deformation of the mechanical structure. This deformation is converted to a measurable electrical signal, which is used to extract the quantity of interest. 

The central idea of the suggested approach is that an additional element is provided allowing to actively tune or even switch off the mechanical signal transferred from the structure to the transducer, Fig.~\ref{fig:concept}(b). While the approach is general, its implementation in a resonant vibration beam accelerometer is considered in the present work. The device's operational scenario is based on using electrostatic rather than direct mechanical transmission of the inertial signal from the proof mass to the vibrating sensing element, allowing controllable ON/OFF switching of the signal and zeroing of the bias, Fig.~\ref{fig:concept}(a). Note that while electrostatic coupling between the resonant sensing element and the proof mass of an accelerometer was previously suggested~\cite{comi2016sensitivity,peng2019sensitivity,Shin2019}, the SF thermal sensitivity prediction and zero bias correction were not addressed. 
The goal of this work is to investigate, using the model, the feasibility of the suggested approach, its main features, limitations and expected performance, along with the evaluation of the envelope of the design parameters toward the realization of the sensor.

The structure of the paper is the following. In Sec.~\ref{sec:design}, we describe the device architecture and operational principle in general. In Sec.~\ref{sec:RO}, we provide the theoretical feasibility study for the proposed approach: Subsec.~\ref{subsec:formulation} gives the formulation of the device's lumped model and a reminder of the definitions for the accelerometer's figures of merits; Subsec.~\ref{subsec:Cap} employs the simplest parallel-planes capacitor model of the electrostatic fields and studies the structure of the device's equilibrium states, their stability and bifurcations, providing the intuition and the analytical estimates for some values of interest; Subsec.~\ref{subsec:FFields} considers the fringing electro-static fields. Sec.~\ref{subsec:Comsol} describes the simplified finite element multiphysics model of the full device that validates the concept numerically. Sec.~\ref{sec:ZAU} provides the insight to the possible technique of self-calibration and temperature compensation. The analytical relationships between the scale factor and the geometric parameters, voltage, acceleration, and the temperature are presented in~\ref{App:lumpedanalytic} and \ref{App:softening}.

\section{Device architecture and operational principle} 
\label{sec:design}
\begin{figure}[h]
    \centering
    \includegraphics[width = 0.85\textwidth]{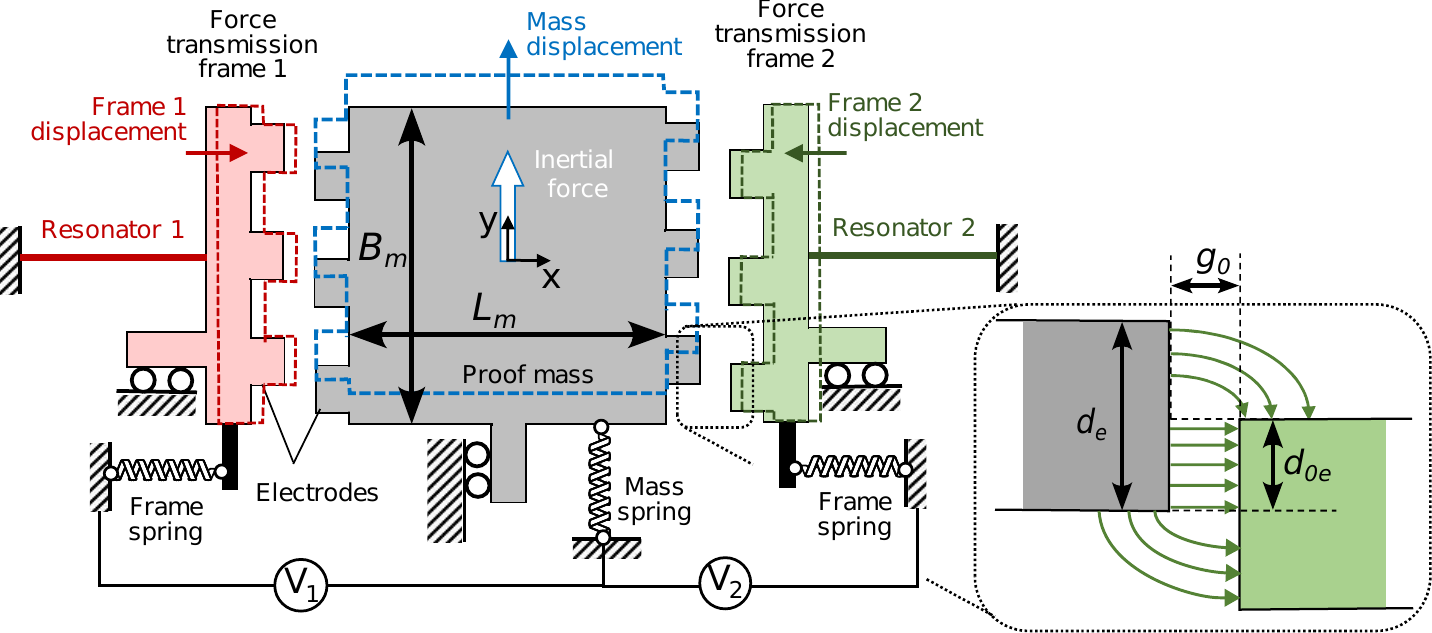}
    \caption{Schematics of a generic differential resonant accelerometer with tunable and switchable force transmission. The dashed lines depict a deformed configuration of the mass and of the frames, the arrows represent the directions of the displacements under an inertial force in the positive $y$ direction. An inset shows an enlarged view of the electrodes and the schematics of the ES field lines (arrows).}
    \label{fig:geometry}
\end{figure}

To highlight  ideas, we first consider the simplest mass-spring representation of the generic device with switchable mechanical signal transmission,  Fig.~\ref{fig:geometry}. The device consists of a proof mass (PM) and  two force-transmitting frames. The mass and the frames are attached to the substrate by suspension springs. Each of the two vibrating sensing beams (the resonators) is attached to the corresponding frame at one end and to the substrate at the other end. The mass is constrained  to move solely in the vertical $y$ (sensing) direction, whereas the frames are designed to move only in the horizontal $x$ direction. Two sets of non-interdigitated comb-like electrodes~\cite{Adams1998172},
operating simultaneously in gap-closing (parallel-plate) and area-changing (comb) modes, are attached to the mass and to the frames. The mass is electrically grounded, and the steady, time-independent, tuning voltages $V_1$ and $V_2$ are applied to the frame $1$ and $2$, respectively. 
Since there is no mechanical interaction between the proof mass and the frames, (perhaps with the exception of possible acoustic coupling through the substrate)  no force is transmitted from the mass to the resonators at zero voltages, irrespective of the proof mass position. Application of the voltages results in the emergence of the ES forces acting on the frames in the $x$ direction in such a way that both resonators are under tension. This tension force is parameterized by the tuning voltages $V_1$ and $V_2$,  and depends on the area of the  electrodes' sidewalls facing each other and on the distance between the electrodes.  The displacement of the mass due to the inertial force results in the disturbance of the ES field and in the change of the force transmitted to the 
resonators. Specifically, due to the configuration of the non-interdigitated electrodes, 
the mass displacement in the positive $y$ direction results in the
increase of the electrode's overlapping area on the left side of the mass (Frame 1),  and decrease on the right side (Frame 2). As a result, the tension of the left resonator $1$ increases, whereas the tension in the right resonator $2$ decreases. By measuring the frequency shift between the resonators,
the acceleration force is extracted. 

Note that the nonlinear ES forces acting on the mass in the $x$-direction can in principle result in the undesirable motion of the mass and so-called pull-in instability; although, the pull-in is prevented by the stiffness of the PM suspension.
Furthermore,  the mass and the frames are subjected to the forces in the $y$-direction  because of the change in the electrode overlap area (``comb-drive'' mode). However, as we show later, the ES forces acting on the mass at the right and the left sides of it are in opposite directions and are compensated to the leading order. The displacement of the frames in the $y$-direction is prevented by the frame suspension design. (Another set of stationary compensation electrodes can be provided to eliminate the influence of the $y$-force acting on the frames.)

The electrostatic coupling between the proof mass and the sensing beams allows to achieve 
two important key features: {\it (i)}  by zeroing the voltages between the mass and the resonators, the influence of the inertial signal can be switched off, and  {\it (ii)} the scale factor of the sensor is parameterized by the coupling voltage and can be tuned in a wide range. 
Note that the active switching or tuning of the  force transmitted from the proof mass to the sensing element can also be achieved by different means by using alternative electrostatic actuating approaches (such as piezoelectric or magnetic) or by incorporating appropriately designed geometrically nonlinear structural elements. In the present work, the considerations are limited to the electrostatic coupling, mainly due to the simplicity of the ES actuation implementation and its widespread use in  MEMS.

\begin{figure}
    \centering
    \includegraphics[width = 0.9\textwidth]{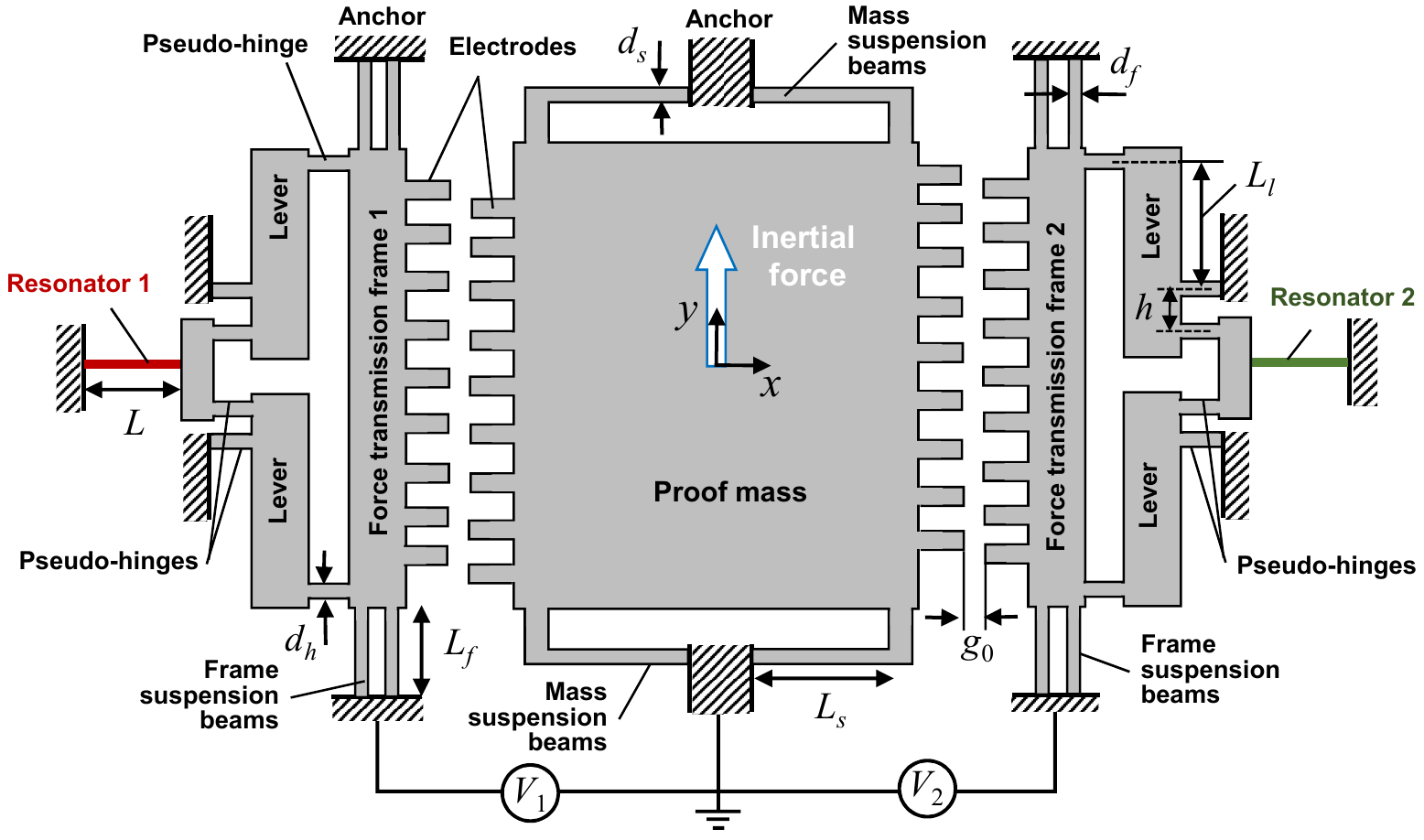}
    \caption{Architecture of the considered resonant accelerometer with lever-type force amplifier and electrostatically  tunable/switchable force transmission.}
    \label{fig:geom_ampl}
\end{figure}

The simplest generic mass-spring model, 
Fig.~\ref{fig:geometry}, does not incorporate force amplification and, while convenient  for the illustration of the suggested  operational principle, cannot be directly implemented  for the feasibility study of the devices of realistic configuration. 
For this reason, a modified, more realistic architecture of the device with an integrated  lever-type compliant force amplification mechanism~\cite{seshia2002vacuum} is considered,
Fig.~\ref{fig:geom_ampl}. In this device, the mass and each of the frames are attached to the substrate by a set of four clamped-guided beams, enforcing the mass and the frames displacement predominantly in the $y$ (sensing) or $x$ directions, respectively.   
The device is assumed to be fabricated of single crystal silicon, using a common deep reactive etching (DRIE)-based process. Similarly to the device shown in Fig.~\ref{fig:geometry}, the main distinguishing element of the suggested architecture is the electrostatic transmission of the inertial force from the proof mass to  the resonant sensing beams, achieved using an array of the non-interdigitated electrodes, attached to the mass and to the force-transmitting frame.

\section{Model} \label{sec:RO}
\subsection{Formulation} \label{subsec:formulation}
The lumped model of the device with force amplification is built under common simplifying assumptions of a perfectly rigid mass, levers and frames and mass-less flexible suspension beams,  pseudo-hinges, and sensing resonators. The equations of equilibrium are first used to evaluate the axial force acting on the sensing beams; the natural frequencies of the resonators are then calculated separately, by considering the double-clamped stretched beam vibrations. 

The equilibrium equations are obtained using the variational principle. The potential energy of the device is the sum of the strain energy of the suspensions, of the hinges, of the resonators (undergoing an axial deformation),  the ES co-energy, and the potential of the inertial force,
\begin{equation}
U=\frac{k_mu^2}{2}+\sum_{i=1}^2 \left[\frac{k_fv_i^2}{2}+\frac{k_h\psi_i^2}{2}+\frac{k_t\Delta L_i^2}{2} -\frac{C_iV_i^2}{2}\right] - ma u
\label{eq:energy_general}
\end{equation}
Here $u$ is the displacements of the mass in the positive $y$ direction, $v_1,\ v_2$  are the displacements of the left and right frame respectively, ($v_i$ is positive toward the proof mass); $\Delta L_i$ is the elongation of the  $i^{\mathrm{th}}$ sensing beam;  $\psi_i$ is the rotation angle of the levers attached to the $i^{\mathrm{th}}$ frame; 
$k_m=48EI_s/L_s^3,\;k_f = 48EI_f/L_f^3 $ is the equivalent stiffness of the mass and the frame suspensions, respectively; $k_h=6EI_h/L_h$ is the combined stiffness of  the six pseudo-hinges attached to the $i^{\textrm{th}}$ frame and $k_t = EA_t/L$ is the axial (tensile) stiffness of each of the resonators. 
Here $E$ is the Young's modulus of Si,  $I_{\{s,f,h\}}=bd_{\{s,f,h\}}^3/12$ are the second moments of the cross-section area of the corresponding elements, $b$ is the thickness of the device, $d_{\{s,f,h\}}$ and $L_{\{s,f,h\}}$ are the width and length of the mass suspensions, frames suspensions and hinges, respectively; $A_t=bd_t$ is the resonator cross-section area.
In addition, $C_i(u,v_i),\;i=1,2$ is the capacitance of each of the sets of electrodes and $a$ is the acceleration. 

The angles of the levers $\psi_i$ are related to the displacements of the frames $v_i$ and the elongations of the resonators $\Delta L_i$ through kinematic relations, that  under the assumption $\psi_i \ll 1$ read
\begin{equation}
    \psi_i \approx \frac{v_i}{L_l},\quad \Delta L_i \approx h  \psi_i = \frac{v_i}{{\cal A}_0}.
\end{equation}
Here $h$ is the lever off-set, $L_l$ is the lever length (see Fig.~\ref{fig:geom_ampl}), and 
\begin{equation}
 {\cal A}_0 = \frac{L_l}{h}
\label{eq:AmplificationA0}
\end{equation}
is the geometric amplification.

By using the principle of stationary potential energy, the equilibrium equations are obtained
\begin{eqnarray}
\label{equi1}
k_m u  &=& F_{y1}+F_{y2}+ma, \\
\label{equi2}
\keff v_i 
&=& F_{xi},\qquad i=1,2.
\end{eqnarray}
Here
\begin{equation}
k_{\mathrm{eff}}
= \frac{k_t}{\mathcal{A}_0^2} \left(1+\frac{\mathcal{A}_0^2\kfs}{k_t}\right),\qquad \kfs = {k_f}+\frac{k_h}{L_l^2},
\label{eq:keff}
\end{equation}
is the effective stiffness of the sensing subsystem (the frame and the resonator) and
\begin{equation}
\label{eq:esforces}
F_{xi}=\left(\frac{\partial C_i}{\partial v_i}\right)\frac{V_i^2}{2},\qquad F_{yi} = \left(\frac{\partial C_i}{\partial u}\right)\frac{V_i^2}{2},\qquad i=1,2
\end{equation}
are the electrostatic forces between the electrodes of each of the sets.
Equations~(\ref{equi1}),~(\ref{equi2}) are the system of three coupled nonlinear algebraic equations in terms of $u,\;v_1,v_2$. The coupling is due to the nonlinear electrostatic forces, which depend on $u,v_1$ and $v_2$.

Once the solution of the system Eqs.~(\ref{equi1}),~(\ref{equi2}) is found, the axial force acting on the resonators can be calculated
\begin{equation}
N_i = k_t\Delta L_i=\frac{k_t v_i}{\mathcal{A}_0}={\mathcal{A}}{F_{xi}}, 
\label{eq:axialforce}
\end{equation}
Here 
\begin{equation}
\mathcal{A}  = \frac{N_1}{F_{x1}}=\frac{k_t}{k_{\mathrm{eff}}\mathcal{A}_0}=
\mathcal{A}_0\left(1+\frac{\mathcal{A}_0^2\kfs}{k_t}\right)^{-1}
\label{eq:amplification}
\end{equation}
is the (non-differential) mechanical amplification considering the signal-transmitting forces $F_{x1}$ as a distinct parameter and evaluating it from Eq.~(\ref{equi2}).
The fundamental mode frequency of the stretched sensing beam is
\begin{equation}
f_i \approx f_0\sqrt{1+\gamma \frac{N_i}{N_E}}
\label{eq:f_0}
\end{equation}
Here $f_0=(\lambda_1^2/2\pi)\sqrt{EI_t/(\rho A_tL^4)}$ is the frequency of the unstretched sensing beam, $\lambda_1 \approx 4.73$ is the eigenvalue of the ideally clamped beam,  $I_t = b d_t^3/12$ is the second moment of the resonator's cross-section, $N_E=4\pi^2 EI_t/L^2$ is the Euler's buckling force and $\gamma \approx 0.96$ is the correction coefficient (hereafter, for the sake of simplicity, this coefficient is omitted). Note that in the actual device, the clamping conditions of the sensing beams are not ideal and the fundamental mode eigenvalue $\lambda_1$ may differ from the theoretical value. In this context, $f_0$ is viewed as the baseline value, which is the resonator's natural frequency in the reference zero-acceleration state, zero voltage and at a reference temperature. 

The resonators' natural frequencies are influenced by the voltage-induced $x$-dis\-place\-ments of the frames. However, in the case of identical resonators and equal applied voltages, the frequencies stay equal for zero inertial input. Thus, in the framework of the differential measurement scenario, the acceleration induces the non-zero frequency shift $\Delta f$ and the scale factor could be defined as
\begin{equation}\label{eq:DeltafSF}
SF =\frac{g \Delta f }{a} = \frac{\Delta f }{\hat{a}} \quad \left(\frac{\textrm{Hz}}{\textrm{g}}\right), \qquad
\Delta f= f_2-f_1 \quad \left(\textrm{Hz}\right),
\end{equation}
where $g = 9.8\ \textrm{m/sec}^2$ is the acceleration of gravity and $\hat{a}=a/g$ is the acceleration expressed in terms of $g$. 
For distinct resonators or different applied voltages, the frequency shift $\Delta f$ should be replaced by a weighted difference of $f_1$ and $f_2$ with the coefficients calibrated for the resonators' geometry and dependent on the voltages \cite{cai2021improved}. In the following, for simplicity and clarity of the development, we study only the case of identical resonators and equal voltages $V_1 = V_2 = V$.

\newcommand{\NL}{{NL}}
In resonant accelerometers, the measurement range of the sensor is often dictated by the acceptable level of the output nonlinearity, which is quantified using the SF nonlinearity
\begin{equation}
NL_{SF} = \frac{SF}{SF_0}-1
\label{eq:SFnonlin}
\end{equation}
where
\begin{equation}
 SF_0 =\lim_{\hat{a} \rightarrow 0} \frac{d\Delta f}{d\hat{a}} \quad \left(\frac{\textrm{Hz}}{\textrm{g}}\right)
 \label{eq:SF0}
\end{equation}
is the (tangent) differential SF at zero acceleration.

In the case of a differential measurement and assuming that the axial forces are much smaller than $N_E$ in Eq.~(\ref{eq:f_0}), the scale factor is 
\begin{equation}
SF = \frac{f_0 mg}{2 N_E}\mathcal{A}_\mathrm{em},\quad \mathcal{A}_\mathrm{em} =\frac{|N_2-N_1|}{ma}.
\label{eq:amplif1}
\end{equation}
Here $\mathcal{A}_\mathrm{em}$ is the electromechanical  transmission force amplification parameter. Note that while 
the mechanical amplification  $\mathcal{A}$  
is dictated solely by the geometry of the lever structure and of the suspensions, the electromechanical amplification parameter $\mathcal{A}_{\textrm{em}}$ also depends on the number and geometry of the electrodes and on the tuning voltage $V$, and can be viewed as a figure of merit of the inertial force to the sensing element transmission.

It is instructive to analyze the mechanical amplification $\mathcal{A}$. Equation~(\ref{eq:amplification}) suggests that in the system incorporating pseudo-hinges and suspensions with finite compliance, the actual mechanical amplification $\mathcal{A}$ differs from (but does not exceed) the geometric amplification $\mathcal{A}_0$. By differentiating $\mathcal{A}$ with respect to the offset $h$ (or $\mathcal{A}_0$), the optimal value of the offset (considered as a design parameter) and the corresponding maximal amplification for the given values of the stiffness $\kfs$ and $L_l$ can be found
\begin{equation}
h_{opt}=L_l\sqrt{\frac{\kfs}{k_t}}, \qquad \mathcal{A}_{\mathrm{max}} =\frac{1}{2} \sqrt{\frac{k_t}{\kfs}}=\frac{L_l}{2h_{opt}}.
\label{eq:optimalh}
\end{equation}
In accordance with Eq.~(\ref{eq:optimalh}), the amplification in a compliant amplifier is equal to or lower than half of the geometric amplification in an ideal amplifier with the optimal offset value~\cite{seshia2002vacuum,Pedersen2004OnAccelerometers}.

\subsection{Electrostatic force - parallel capacitor approximation}
In general,  for the adopted electrodes geometry, the nonlinear electrostatic force is influenced by fringing electrostatic fields, cannot be described by a simple expression and can be evaluated only numerically. 
However, it is instructive to  
consider the simplest parallel-capacitor approximation of  this force. The model is convenient for the preliminary, both qualitative and quantitative, investigation of feasibility of the suggested approach  and its potential performance.

The capacitance of each of the sets of the electrodes is 
\begin{equation}\label{eq:capacitance}
C_{i}(u, v_i)=\frac{n \epsilon_0 b \, \left[d_{0e} +(-1)^{i-1} u\right] }{2 \left(g_0-v_i\right) },\ i=1,2.
\end{equation}
Here
$\epsilon_0=8.854\times10^{-12}$ F/m is  the free space permittivity, $b$ is the thickness of the device layer, and $n$ is the number of the electrodes, $d_e$ is the width of the electrode, $d_{0e} = d_e/2$ is the initial overlap between the electrodes faces, $g_0$ is the initial distance between the electrodes; 
$i=1$ and $i=2$ correspond to the left and the right sets of the electrodes, respectively, Fig.~\ref{fig:geometry}.
Note that, in accordance with
Eq.~(\ref{eq:capacitance}),  due to the motion of the mass and the frames in the $y$ and the $x$ directions, respectively, the relative motion between the non-interdigitated electrodes is two-dimensional and incorporates both $u$ and $v$.

Substitution of Eq.~(\ref{eq:capacitance}) into Eqs.~(\ref{eq:esforces}) and then into the equilibrium Eqs. (\ref{equi1}),  (\ref{equi2}) yields
\begin{equation}\label{equi1cap}
- \frac{n\epsilon_0 b  V^2 }{2 \left( g_0-v_1\right) }+ \frac{n \epsilon_0 b  V^2 }{2 \left( g_0-v_2\right) }
+k_m u- m a= 0,
\end{equation}
\begin{equation}\label{equi2cap}
\keff v_i = \frac{ n\epsilon_0 b  V^{2} \, \left[ d_{0e}+(-1)^{i-1}u\right] }{2 {{\left(g_0 - v_i\right) }^{2}}},
\quad i=1,2.
\end{equation}
It is convenient to re-write  Eqs.~(\ref{equi1cap}),~(\ref{equi2cap}) in the non-dimensional form
\begin{equation}\label{equi1cap_ud}
\hat{u} -\eta\beta\left[\frac{1}{1-\hat{v}_1} - \frac{1}{1-\hat{v}_2}\right]= \tilde{a},
\end{equation}
\begin{equation}\label{equi2cap_ud}
\hat{v}_i = \frac{\beta \left[1+(-1)^{i-1}\hat{u}\right]}{(1-\hat{v}_i)^2},\quad i=1,2,
\end{equation}
where
\begin{equation}
    \hat{u} = \frac{u}{d_{0e}},\ \hat{v}_i = \frac{v_i}{g_0},\quad \eta = \frac{\keff g_0^2}{k_m d_{0e}^2},\ \tilde{a} = \frac{ma}{k_m d_{0e}},\ \beta = \frac{n\epsilon_0 b  d_{0e}}{2 \keff g_0^3 }\,V^2.
    \label{eq: dimensionless_pars}
\end{equation}
Equations~(\ref{equi1cap_ud}) and~(\ref{equi2cap_ud}) incorporate three parameters: the inertial  loading parameter $\tilde{a}$,  the electrostatic loading parameter  $\beta$, and the stiffness ratio parameter $\eta$, which is dictated solely by the geometry.

\begin{figure}
    \centering
    \includegraphics[align=c,width = 0.45\textwidth]{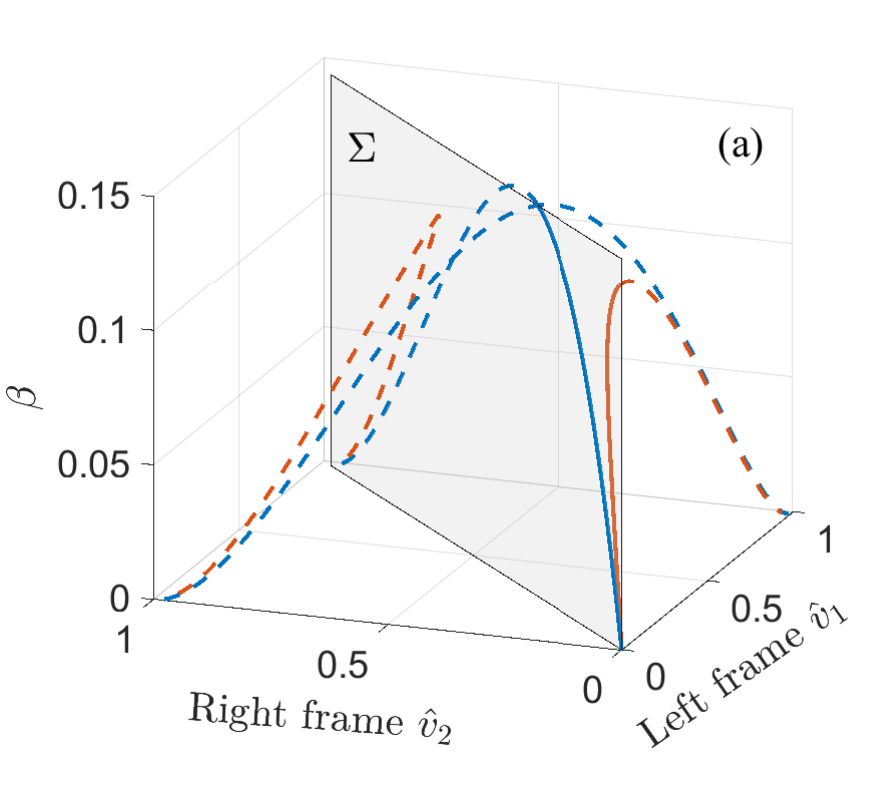}
    \includegraphics[align=c,width = 0.42\textwidth]{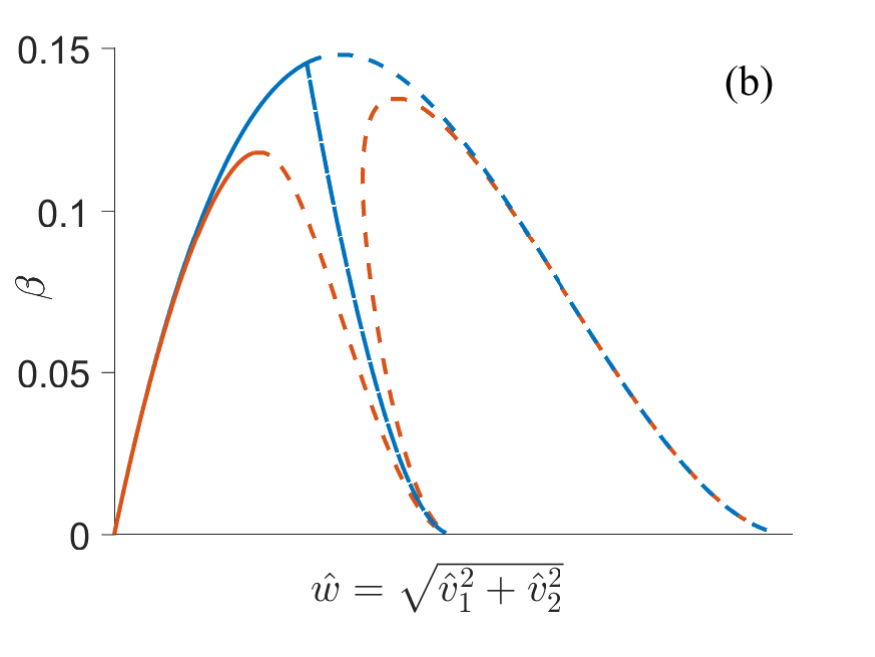}
    \caption{
    (a) Three-dimensional rendering of the equilibrium curve (bifurcation diagram): voltage parameter $\beta$ versus the frames displacements $\hat{v}_1,\;\hat{v}_2$ is shown. The plane $\Sigma$ (shown in gray) 
    corresponds to the symmetric solutions $\hat{v}_1 =\hat{v}_2,\;\hat{u}=0 $. 
    (b) Projection of the equilibrium curves on the plane $\Sigma$. The blue and orange lines on (a) and (b) correspond to $\tilde{a} = 0$ and $\tilde{a} = 0.2$, respectively. Solid and dashed lines correspond to the stable and unstable solutions, respectively.
   }
    \label{fig:equilibrium_curves_cap2}
\end{figure}

\begin{figure}[h]
    \centering
    \includegraphics[width = 0.45\textwidth]{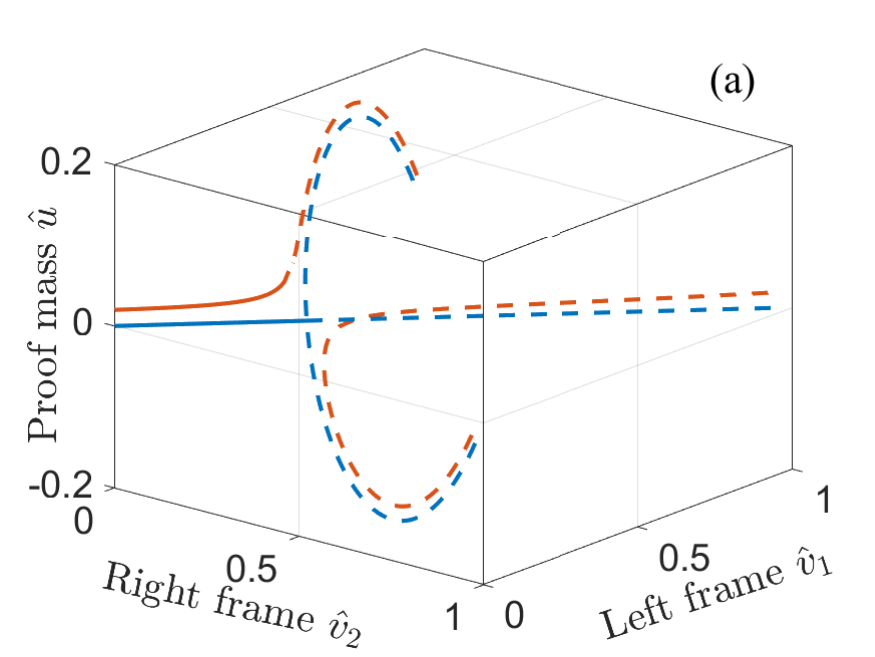}
    \includegraphics[width = 0.45\textwidth]{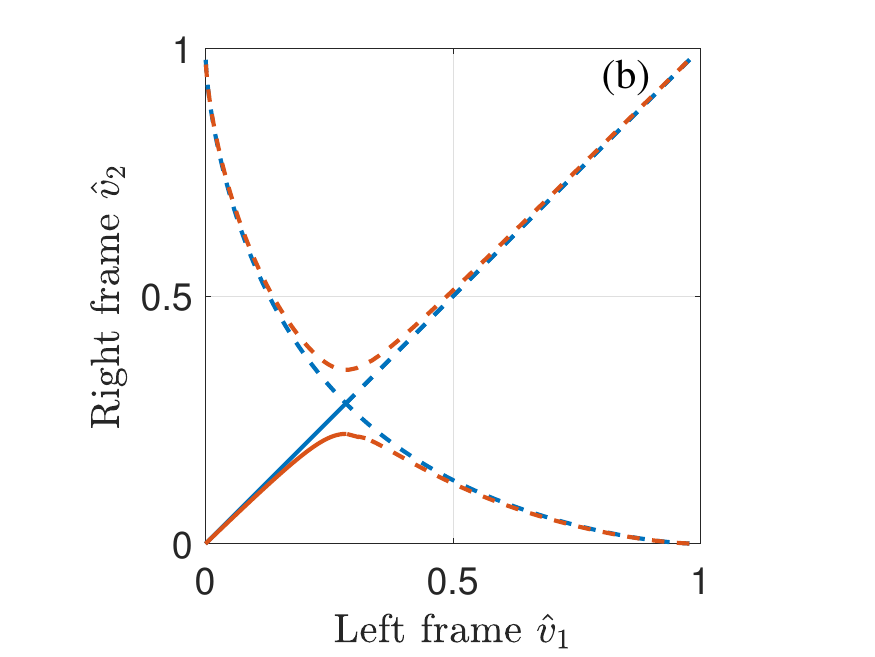}
    \caption{(a) Three-dimensional rendering of the equilibrium paths: proof mass displacements $\hat{u}$ versus frames displacements $\hat{v}_1$, $\hat{v}_2$, for $\tilde{a} = 0$ (blue lines) and $\tilde{a} = 0.02$ (orange lines). (b) Projection of the equilibrium path on the $\hat{v}_1,\;\hat{v}_2$ plane. Solid and dashed lines on (a),(b) correspond to stable and unstable branches, respectively. 
    }
    \label{fig:equilibrium_curves_cap1}
\end{figure}

{\bf The case of zero acceleration.} At $\tilde{a} = 0$ there exists a symmetric solution  $\hat{u}=0$, $\hat{v}_1 = \hat{v}_2$. This case is similar to the well-known canonical spring-capacitor model \cite{senturia2005microsystem}. The equilibrium curve (the bifurcation diagram) is located within the plane   $\hat{v}_1 = \hat{v}_2$ and contains a stable branch corresponding to the range $\hat{v}_{1,2} < 1/3$ and an unstable branch for $\hat{v}_{1,2} \geq 1/3$. The critical (pull-in) point separating the branches corresponds to  $\beta_{PI} = {4}/{27}$.
For the device design parameters listed in Tab.~\ref{tab:values} (which are used to obtain all the numerical results presented hereafter), the corresponding critical value of the voltage is $ V_{PI} = \sqrt{8\,\keff\,g_0^3/(27\,n \epsilon_0 b \, d_{0e})}\approx 62.5$ V.
Figures~\ref{fig:equilibrium_curves_cap2}(a) and (b) depict the three-dimensional rendering of the bifurcation diagram and its projection to the diagonal plane $\Sigma$ corresponding to $\hat{v}_1 = \hat{v}_2$. The equilibrium path within the $\hat{u},\;\hat{v}_1,\;\hat{v}_2$ space and its projection on the $(\hat{v}_1,\;\hat{v}_2)$ plane are shown in Fig.~\ref{fig:equilibrium_curves_cap1}(a), (b), respectively. 

Non-symmetric solutions $\hat{u} \neq 0,\;\hat{v}_1 \neq \hat{v}_2$ at $\tilde{a}=0$ include two unstable branches (shown by dashed blue lines in Figs.~\ref{fig:equilibrium_curves_cap2},~\ref{fig:equilibrium_curves_cap1}), emerging from the symmetric branch through subcritical pitchfork bifurcation. The location of the bifurcation point on the symmetric branch is dictated only by the stiffness ratio parameter $\eta$. The symmetric branch $\hat{u}=0$ in Fig.~\ref{fig:equilibrium_curves_cap2} is stable  up to the pitchfork bifurcation point. For the parameters listed in Tab.~\ref{tab:values},  $\eta = 1.3$ and the pitchfork bifurcation occurs at $\beta = 0.145 <  \beta_{PI}$,  (that corresponds to $V_{b}\approx 62$~V). Increase of $\eta$ (which corresponds to larger effective stiffness $k_{\textrm{eff}}$ with respect to the mass suspension stiffness $k_m$, Eq.~(\ref{eq: dimensionless_pars})) results in bifurcation occurring at smaller displacements and smaller $\beta$.  In contrast, at $\eta \rightarrow 0$, the pitchfork point approaches (but does not surpass) the symmetric pull-in point $\hat{v}_1=\hat{v}_2=1/3,\;\beta_{PI}=4/27$.  

\begin{table}[htbp]
\caption{\textbf{Parameters of the device used in the calculations}
}
    \centering
    \begin{tabular}{|l|c|c|}
    \hline
         \textbf{Parameter}  & \textbf{Unit} & \textbf{Value}\\
         \hline
        Density $\rho$ & kg/m$^3$ & 2300
         \\
         \hline
         Young's modulus $E$ & GPa & 169
         \\
              \hline
        Thickness in the $z$-direction $b$& $\mu$m  & 50
         \\
         \hline
         Proof mass size $L_m\times B_m$& $\mu$m  & 2000$\times$2500
         \\
         \hline
           Pseudo hinge length and width $L_h \times d_h$ & $\mu$m &	32 $\times$ 8 
         \\
         \hline
            Sensing beam length and width  $L \times d$ 	& $\mu$m	&  300 $\times$ 4
        \\
         \hline
         Mass suspension beam length and width $L_s \times d_s$ & 	$\mu$m	& 450 $\times$ 6
        \\
         \hline
          Frame suspension beam length and width $ L_f \times d_f$	&  $\mu$m& 	900 
        $\times$ 6
        \\
         \hline
        Lever length $L_l$	&$\mu$m &	870
        \\
         \hline
               Offset $h$ & 	$\mu$m	&  30
        \\
         \hline
        
        Gap $g_0$ & 	$\mu$m	&  2.5
        \\
         \hline
        
        Electrode length  and width  $L_e\times d_e$ & 	$\mu$m	 & 17$\times$8
        \\
         \hline
         Pitch between the electrodes $p_e$ & 	$\mu$m	 & 16
        \\
         \hline
                Number of electrodes $n = \lfloor{B_m}/{p_e}\rfloor$ (one side) & --&		156
        \\
         \hline
            \end{tabular}
    \label{tab:values}
\end{table}

\begin{figure}[h]
    \centering
    \includegraphics[width = 0.45\textwidth]{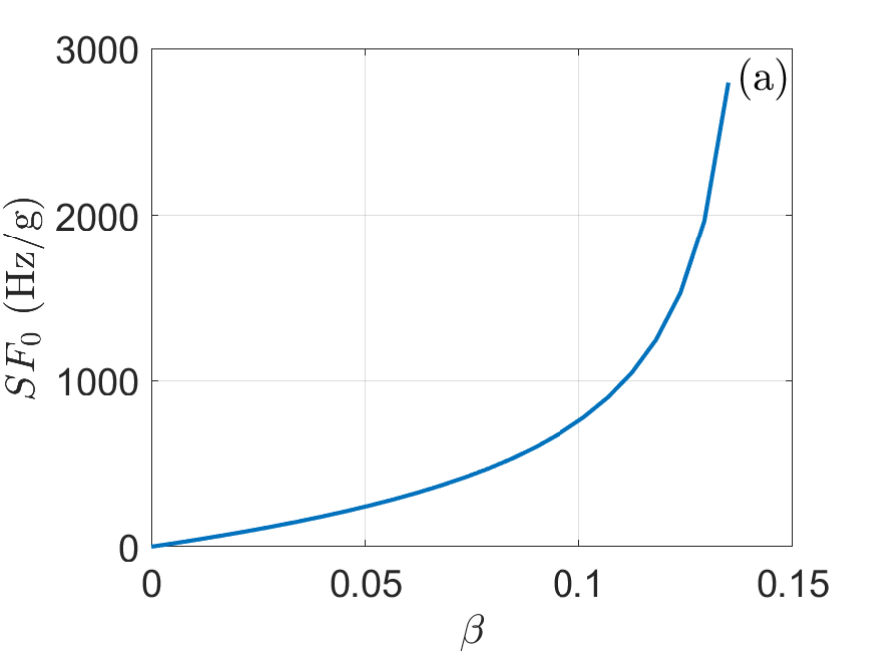}
   \includegraphics[width = 0.45\textwidth]{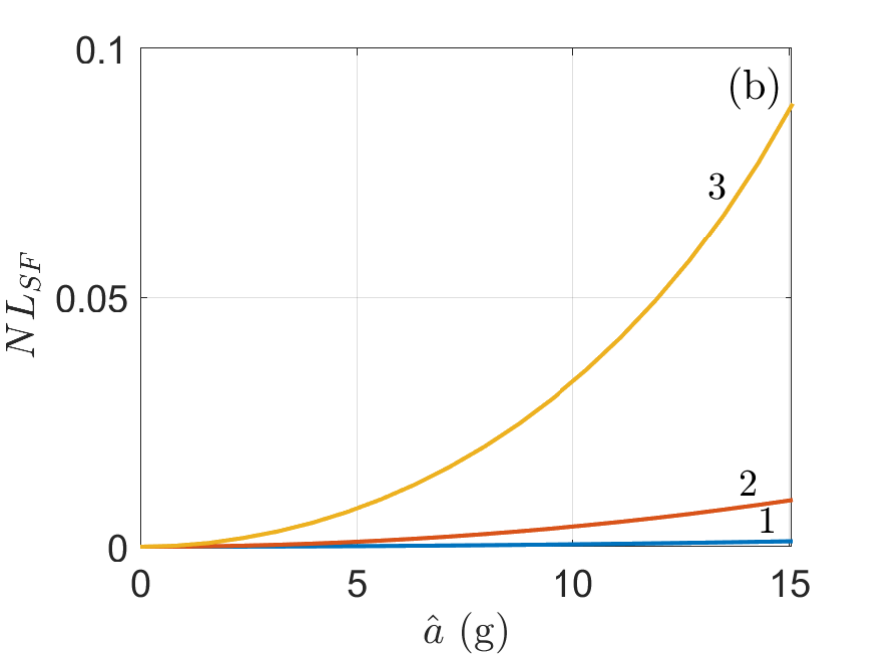}
    \caption{(a) Tangent differential scale factor $SF_0$ (Eq.~(\ref{eq:SF0})) versus $\beta$. The critical pull-in value of the voltage parameters is $\beta_{PI} =4/27 \approx 0.148$. (b) Scale factor nonlinearity $NL_{SF}$ 
    for several values of $\beta = 0.033$ (line 1), $\beta = 0.067$ (line 2), $\beta = 0.1$ (line 3) as a function of the acceleration $\hat{a}$.
    }
    \label{fig:SF_cap}
\end{figure}

{\bf The case of nonzero acceleration.} In this case, $\hat{u} \neq 0$ (see Eq.~(\ref{equi1cap_ud}))
and the displacements of the frames $\hat{v}_1 \neq \hat{v}_2$ (see Eqs.~(\ref{equi2cap_ud})) and consequently the axial forces,  Eq.~(\ref{eq:axialforce}), are not equal, which makes possible the differential measurement of the acceleration.
As the acceleration $\tilde{a}$ varies, the pitchfork splits into two distinct branches of solutions  (see Fig.~\ref{fig:equilibrium_curves_cap2}, Fig.~\ref{fig:equilibrium_curves_cap1}, orange lines).
The resulting scale tangent differential scale factor $SF_0$ and its nonlinearity $NL_{SF}$ ( (Eqs.~(\ref{eq:SFnonlin}), (\ref{eq:SF0})) are shown in Fig.~\ref{fig:SF_cap}. One observes that, first, as $\beta$ approaches the critical value, the scale factor grows fast, indicating that tuning by the  applied voltage is possible. Secondly,  nonlinearity of the scale factor also increases with~$\beta$.
\label{subsec:Cap}
\subsection{Electrostatic force -- numerical model} \label{subsec:FFields}
\newcommand{\Eshape}{\rotatebox[origin=c]{-180}{${\exists}$}}
\begin{figure}
    \centering
    \includegraphics[scale=.45]{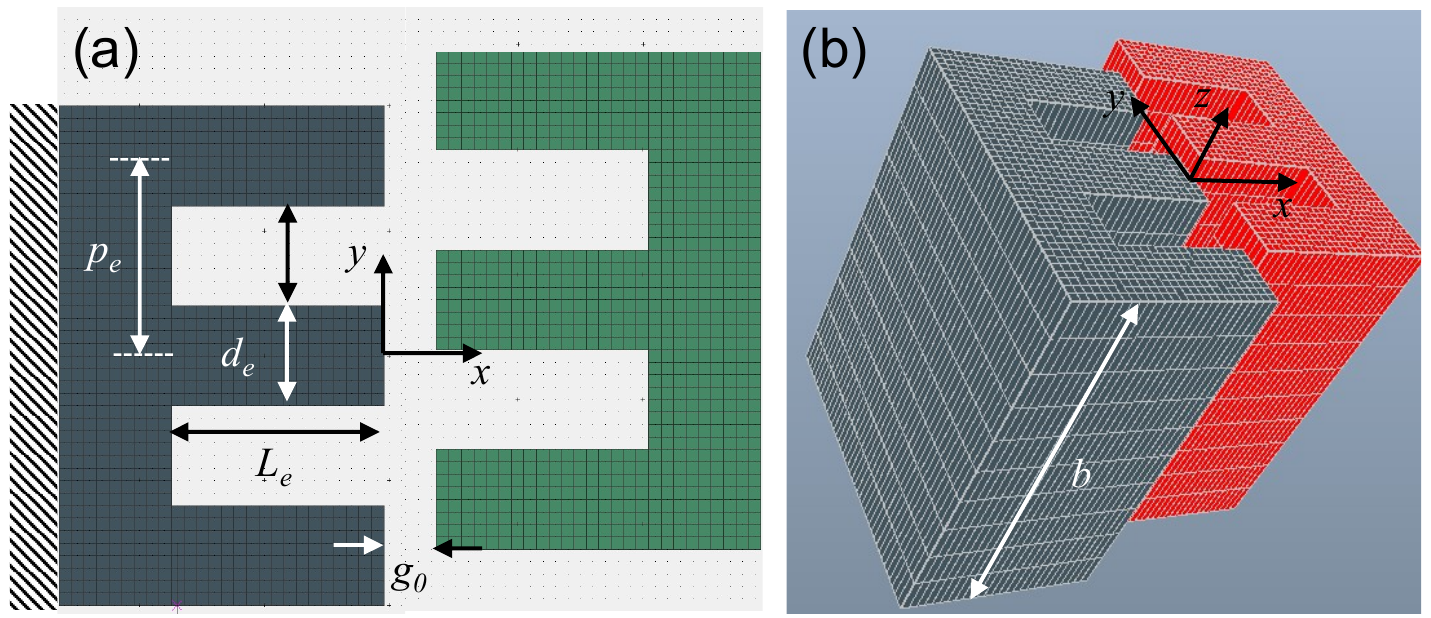}
    \caption{Electrostatic forces evaluation - a model of the three-finger electrode arrangement and the Boundary Element Method mesh (Intellisuite) in the initial configuration. (a) - top view (b) isometric view.}
    \label{Fig:IntellisuiteThree}
\end{figure}

\begin{figure}
    \centering
    \includegraphics[scale=0.5]{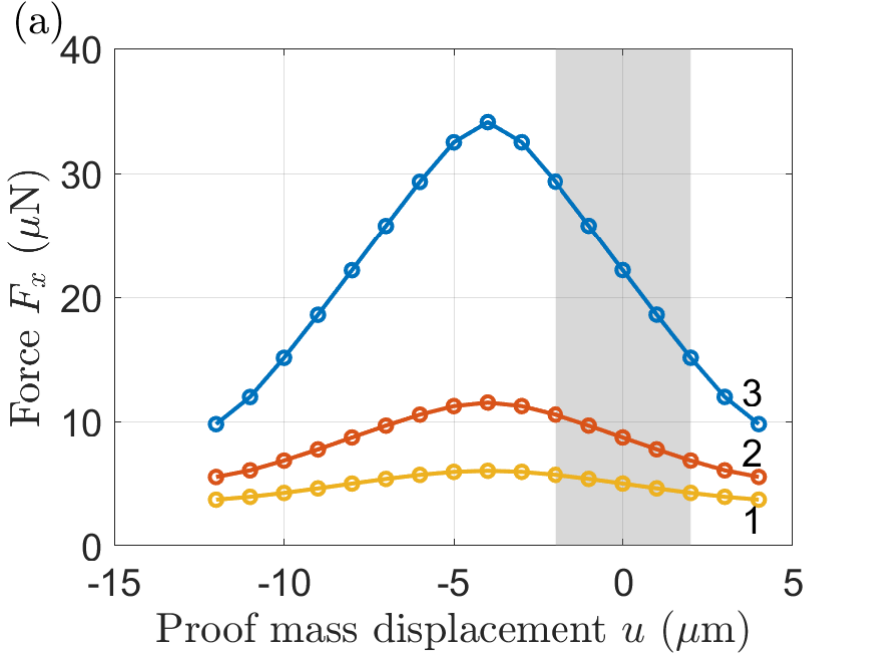}
    \includegraphics[scale=0.5]{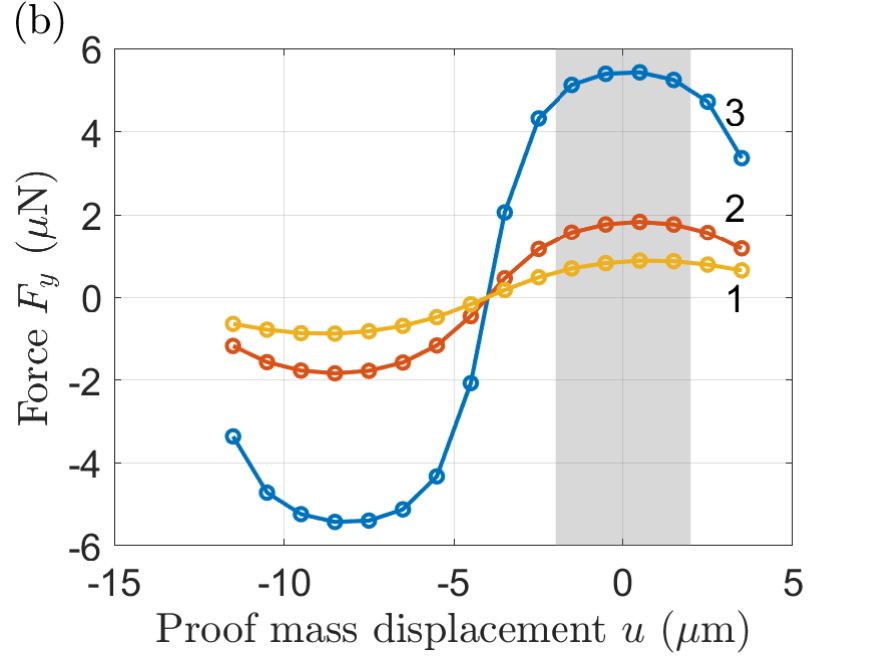}
    \caption{BEM numerical results: electrostatic force $F_x$ (a) and $F_y$ (b) in the lateral and the sensing directions, respectively, as a function of the relative displacement $u$ between the electrodes in the case of the three-electrodes configuration for several values of the frame displacement $v$ towards the mass. The voltage is $V=$100 V, the gap $g_0 = 2.5\,\mu$m. Gray rectangles represent the working regions $u\in[-2,2]\, (\mu \textrm{m})$. 
  {(a): $v = 0\,\mu$m (line 1), $1\,\mu$m (line 2), $2\,\mu$m (line 3); (b): $v = -0.5\,\mu$m (line 1), $0.5\,\mu$m (line 2), $1.5\,\mu$m (line~3).}}
    \label{Fig:ESforces_three}
\end{figure}

The simple parallel-plate capacitor approximation does not take into account the influence of the fringing electrostatic fields appearing due to the finite electrodes width $d_0$, which is comparable to the gap $g_0$ between the electrodes. In the framework of the more realistic model, the capacitance between the two sets of the electrodes and the electrostatic forces $F_{xi}$, $F_{yi}$  were calculated numerically by the boundary elements method (BEM), using the IntelliSuite package~\cite{IntelliSuite}. 
In order to increase the interaction force between the electrodes and reduce the required voltage, the actual design incorporates numerous ($n=156$, see Tab.~\ref{tab:values}) electrodes, Fig.~\ref{fig:geom_ampl}. Direct evaluation of the ES force for such a large number of electrodes is computationally very intensive. For this reason, calculations were carried out for the sets containing just three brick-shaped electrodes (``fingers"), as shown in Fig.~\ref{Fig:IntellisuiteThree}. The force provided by the $n$-electrodes set was then re-scaled by multiplying the results by $n/3$ (under the assumption that the ES field distribution is periodic within the set). The dimensions of the fingers are $L_e\times d_e\times b = $17  $\mu$m $\times$ 8  $\mu$m  $\times$ 50  $\mu$m  and there is a pitch of $p_e = $16  $\mu$m  between them.  The fingers are attached to a brick-shaped element, with the dimensions of 9 $\mu$m $\times$ 40 $\mu$m $\times$ 50 $\mu$m representing a part of the frame or of the proof mass. Here in the initial reference configuration (for $u = 0\ \mu$m), the overlap  between the mass and the frame electrodes is $d_{0e} = d_e/2  = 4\, \mu$m.
The dependence of the electrostatic forces on the mass displacement~$u$ for several values of the displacement $v$ in the $x$ direction are depicted in Fig.~\ref{Fig:ESforces_three}. 
As expected, when the electrodes are fully aligned $u = -4\ \mu$m, the attractive force  $F_{x}$ is maximal, Fig.~\ref{Fig:ESforces_three}(a), and the forces $F_{y}$ in the sensing direction is zero, Fig.~\ref{Fig:ESforces_three}(b); $F_{y}$ is maximal when $u = 0$ (the initial reference configuration).  Hence,  the working region can be found such that the dependence of $F_x$ on the displacement $u$ is close to linear and the sensitivity of $F_x$ to $u$ (the slope of the $F_x(u)$ curve) is maximal. In contrast, the force $F_y$ is close to constant there, and the sensitivity of $F_y$ to the relative displacement is minimal. 

We approximate the capacitance in the working range by a 4$^{\mathrm{th}}$ order polynomial  in $u$ and 2$^{\mathrm{nd}}$ order in $v$:\
\begin{eqnarray}\label{eq:fit_NL}
    C_{el}(u,v_i) &=& \sum\limits_{r=1}^{5}\sum\limits_{s=1}^{3} \tilde{c}_{r,s}u^{5-r}v_i^{3-s},
\end{eqnarray}
where the coefficients were adjusted to fit the numerical data as
\begin{equation}
    \tilde{c}_{r,s} =10^{-5}\times 
\begin{pmatrix}
 1.9 & 0.82 & -0.15\\
\mp 7.4 &  \mp 10.5 & \mp 16 \\
16 &  25 & 16 \\
\pm 1359&   \pm 977 & \pm 1062 \\
5634 &     8012 &       42805\\
  \end{pmatrix}\label{eq:fit_coef}	
\end{equation}
Hereafter, the upper sign stands for the left frame $v_1$, and the lower sign for the right frame $v_2$; measurement units of the coefficients are not the same and correspond to the numerical value of $C_{el}(u,v_i)$ in pF if the displacements $u$ and $v_i$ are  in $\mu$m. The results of the numerical experiments conducted with one-, three-, and five-finger electrodes indicate that the dependence of the mutual capacitance's between the electrode's sets on the number of the fingers is close to linear and the re-scaling of the force is justified. 
Therefore, considering the number of electrodes (Tab.~\ref{tab:values}), the capacitance coefficients $\tilde{c}_{r,s}$, Eq.~(\ref{eq:fit_NL}) are proportional to the numerical values, obtained by fitting of the data provided by the BEM  analysis,  with the factor of 156/3.
In view of the polynomial approximation Eq.~(\ref{eq:fit_NL}) of the capacitance associated with fringing electrostatic field,  Eqs.~(\ref{equi1}), (\ref{equi2}) are reduced to the three polynomial equations in terms of $u$, $v_1$ and $v_2$.
Their solutions were obtained numerically for the parameters of the device listed in Tab.~\ref{tab:values} and the fitting coefficients from Eq.~(\ref{eq:fit_coef}).

\begin{figure}
    \centering
    \includegraphics[width = 0.45\textwidth]{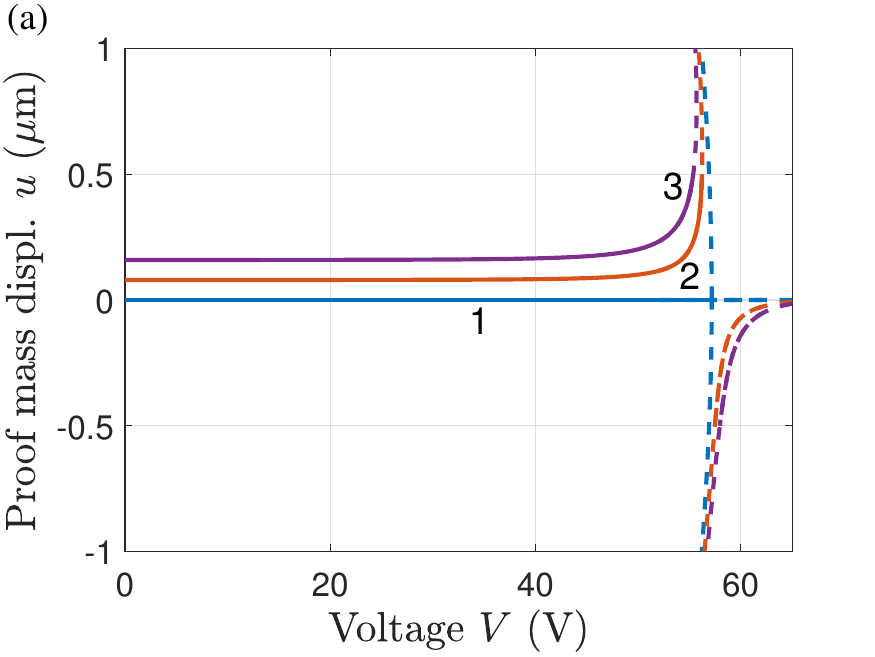}
    \includegraphics[width = 0.45\textwidth]{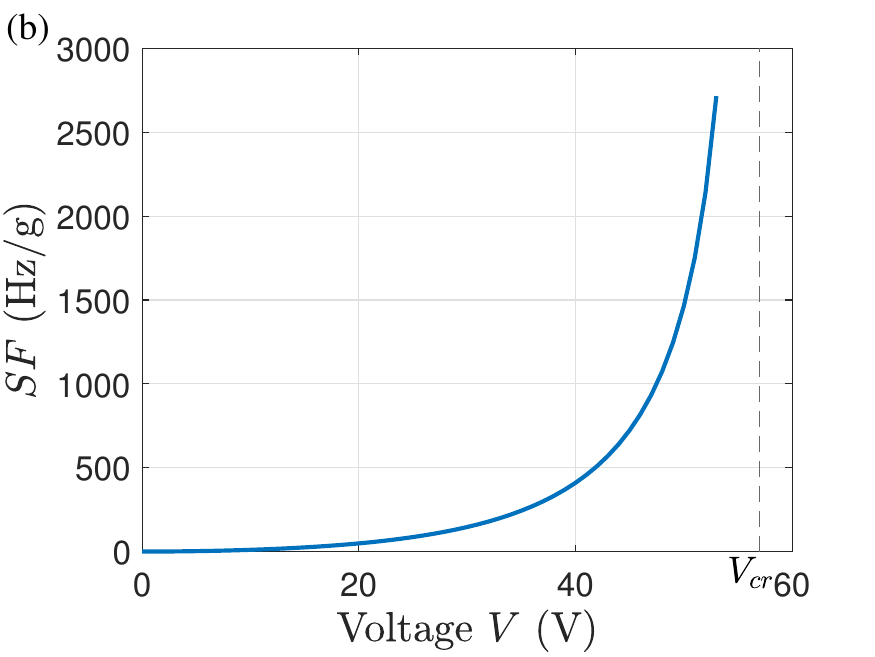}
    \caption{(a): Proof mass displacement $u$ for  several values of acceleration $\hat{a} = 0$ (line 1), $\hat{a} = 1$  (line 2), $\hat{a} = 2$ (line 3) as a function of the voltage $V$. (b): Scale factor versus voltage $V$ at $\hat{a} = 1$, the critical voltage $V_\mathrm{cr}$ is shown by a dotted line.} 
    \label{fig:equilibrium_curves}
\end{figure}

Qualitatively, the behaviour of the system  is similar to that of the parallel  capacitor model described in Sec.~\ref{subsec:Cap}. For zero acceleration $a = 0$,  $v_1=v_2=v_*(V)$ and the trivial equilibrium $u = 0$ is stable up to $V_{\mathrm{cr}} \approx $ 57 (V), where it loses stability through pitchfork bifurcation (see Fig.~\ref{fig:equilibrium_curves}(a), where  the bifurcation diagram is presented only within the stable working range of the device). This critical voltage value is slightly smaller than the pull-in voltage $V_{PI} \approx 62.5$ V predicted by the parallel capacitor model. 
For $a\neq 0$, the pitchfork splits on two separate branches of solutions. The scale factor of the device is shown in Fig.~\ref{fig:equilibrium_curves}b: the SF grows to infinity when the voltage approaches the critical value.%

\begin{figure}
    \centering
    \includegraphics[width=0.45\textwidth]{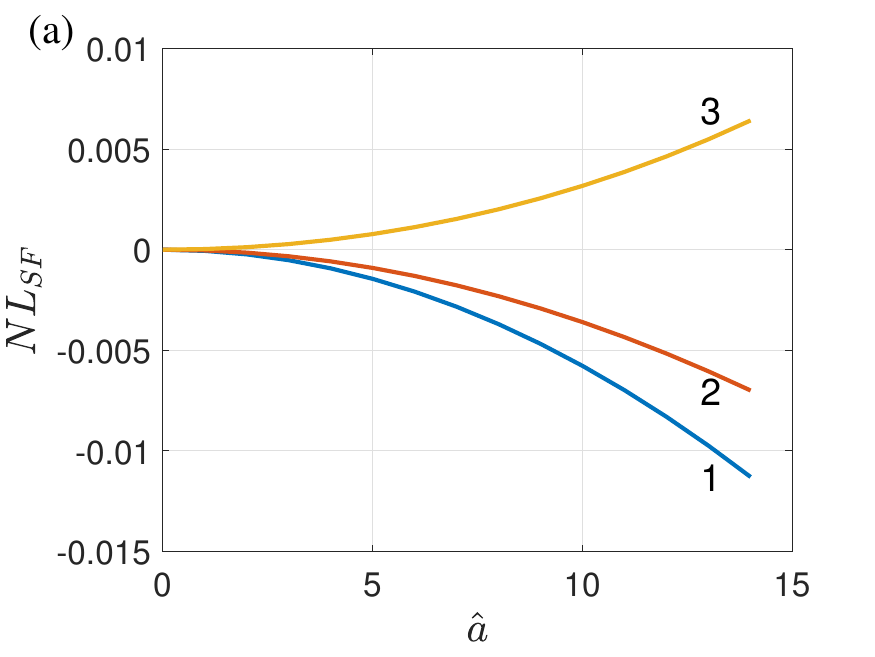}
     \includegraphics[width=0.45\textwidth]{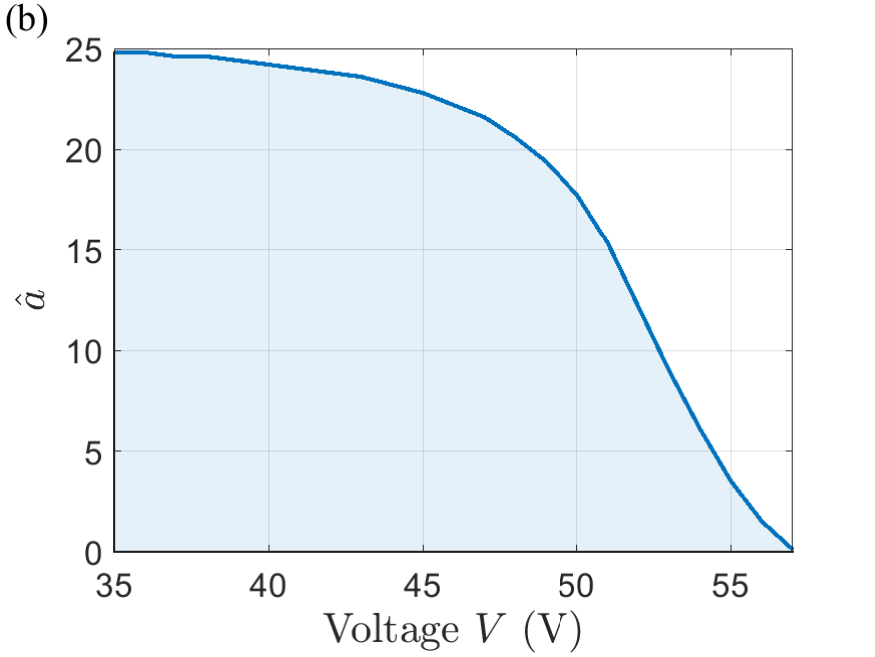}
    \caption{(a) Nonlinearity of the SF 
    versus applied acceleration for several values of voltages $V = 20$ V (line 1), $V = 30$ V (line 2), $V = 40$ V (line 3). (b) Acceleration's working region versus voltage.
    }
    \label{fig:NonlinearityWR}
\end{figure}

Our results indicate that the working range, in terms of the acceleration, of the device considered in the present work is limited either by the frame's pull-in instability or by the nonlinearity of the scale factor (defined in Eq.~(\ref{eq:SFnonlin})).
Both limitations are influenced by the tuning voltage $V$. Namely, application of the higher voltage between the proof mass and the frames results in higher SF, but also causes the device to be more prone to pull-in, and increases the SF nonlinearity.
In addition, the sensor exhibits non-monotonic dependence of  the scale factor non-linearity $NL_{SF}$ on the applied voltage (see Fig.~\ref{fig:NonlinearityWR}(a)). This behavior may open a possibility to the SF nonlinearity tailoring by choosing the suitable tuning voltage. 
Moreover, we can argue that by optimising the electrodes' geometry, it would be possible to overcome, or at least hinder, the SF and SF nonlinearity trade-off 
(the electrodes geometry 
optimization for performance is not considered in the present work).

The dependence of the maximal allowable acceleration on the applied voltage is shown in Fig.~\ref{fig:NonlinearityWR}(b).  Calculations show that for the voltages $V < 51\,$(V), the acceleration input is limited by the proof mass's maximal deflection (it was set at $2\,\mu$m {based} on the fit range for electrostatic forces shown in Fig.~\ref{Fig:ESforces_three}). For higher voltages up to the critical $V_{cr}$, the acceleration limit is bounded by the existence and stability of the working regime.

\section{Finite element analysis} \label{subsec:Comsol}
\begin{figure}
    \centering
    \includegraphics[scale=0.2]%
    {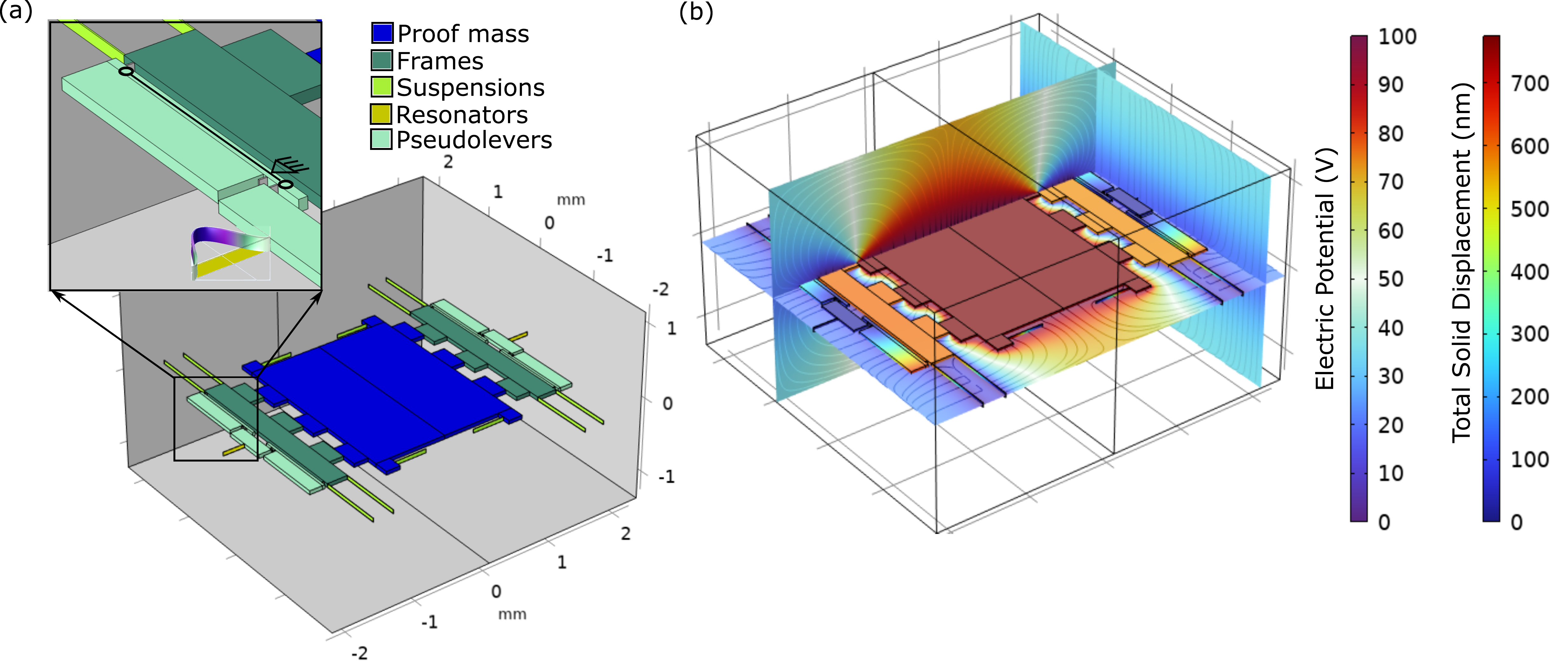}
    \caption{(a) Simplified FEM multi-physics model of the device with three 
    electrodes on the frames and the mass. (b) Results of simulations for $V = 100$ (V) and $\hat{a} = 10$. Electric potential is shown on the plane slices; total solid displacements are shown on solid surfaces.}
    \label{fig:FEMModel}
\end{figure}

To illustrate the feasibility of the suggested SF tuning and calibration approach, we analyze numerically the fully coupled electromechanical finite element (FE) model of the device shown in Fig.~\ref{fig:FEMModel}, with a small number of electrodes and for the design parameters listed in Tab.~\ref{tab:values}, except for the offset that was set to $h = 60$ $\mu$m. Since the transmitting electrostatic forces $F_{xi}$ are proportional to the product $n V^2$, the re-scaling coefficient of the voltage in the lumped model to the voltage in the FE model, which incorporates only three rather than 156 electrodes,  is $\sqrt{156/3}$. We emphasize that the SF of this device is significantly reduced and not realistic. The model is used to show qualitatively the feasibility of the suggested approach.

The model  was built using Comsol Multiphysics Version 6.2 \cite{Comsol} and involves two modules -- solid mechanics  and electrostatics modules, combined by Electromechanical Forces coupled interface (Fig.~\ref{fig:FEMModel}). The device is assumed to be made of a linearly elastic material with the elastic modulus of $E=169$ GPa and Poisson's ratio $\nu=0.06$. Since all the slender deformable structural elements (the mass and the frames suspension beams, the hinges and the sensing resonators) are oriented in the $\langle$ 110 $\rangle$ direction of the Si wafer, and are essentially in a one-dimensional stress state, this assumption is justified \cite{Hopcroft2010WhatSilicon}. To model the electrostatic force taking into account the fringing fields, the height of the electrostatic computational domain of 2.5 mm was chosen in accordance with the recommendation provided in~\cite{comsol_capacitance}. The resonators are attached to the frames by two symmetrical pseudolevers (shown
in the inset in Fig.~\ref{fig:FEMModel}). 
The frames and the mass are attached to the substrate by four beam-type suspensions each. 

The resonators' working frequencies in the reference stress-free configuration were found to be 397628 Hz (the corresponding  mode is shown in the inset of Fig.~\ref{fig:FEMModel}). This value is 2\% higher than the theoretical prediction for the ideally clamped double clamped beam with the same dimensions. The lowest eigenmode of the device is the in-plane  proof mass's translational movement along the $y$-axis with the frequency of 1792 Hz.
 
\begin{figure}
    \centering
    \includegraphics[scale = 0.45]{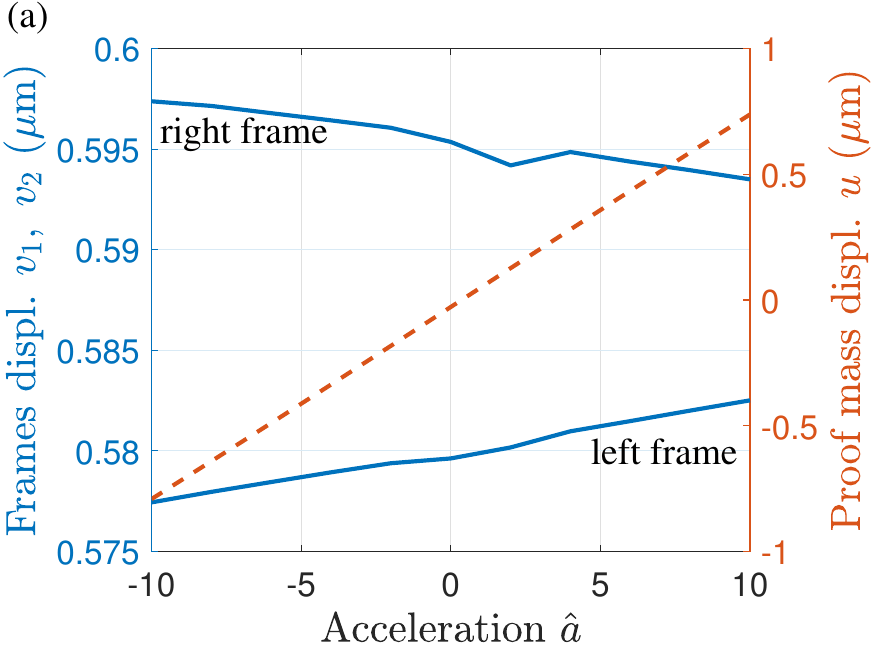}
    \includegraphics[scale = 0.45]{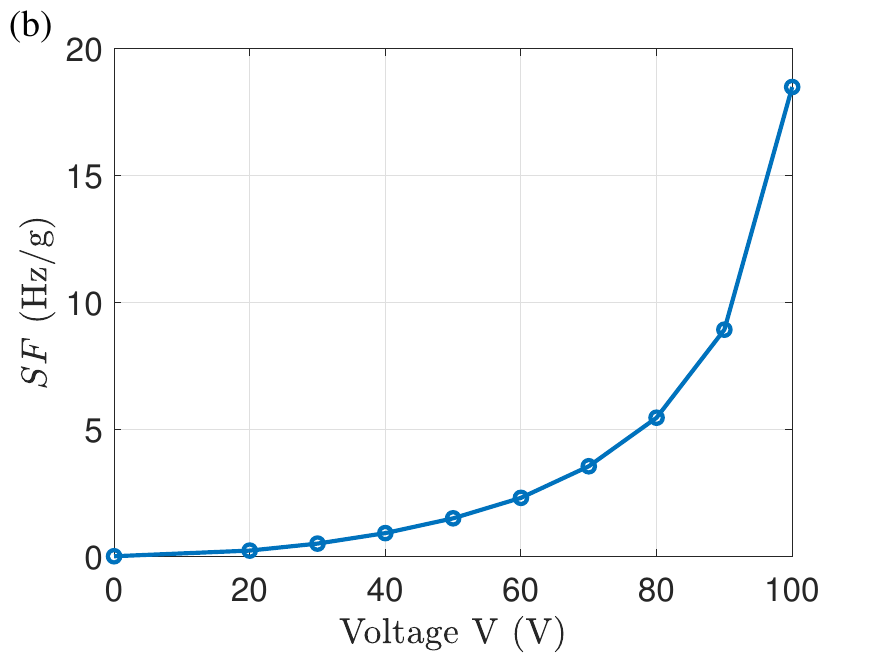}
    \caption{(a): Proof mass (dashed red line) and frames displacements (blue lines) versus acceleration for $V = 100\ (V)$  in FE model. (b) Dependence of the scale factor
    on the voltage.}
    \label{fig:FEMDisplacements}
\end{figure}

The model shows that the electrostatic force can be used to transmit the sensing input (acceleration) to the output signal: the inertial body load along the $y$-axis results in the proof mass and frames' displacements and causes the shift in frequencies. We apply voltage $V = 100$ V. The proof mass and the frames displacements linearly depend on the applied acceleration (Fig.~\ref{fig:FEMDisplacements}a). However, the displacements of the frames for zero acceleration are not equal -- the mismatch is  15.7 nm which appears due to the differences in the mesh around the electrodes and finite numerical accuracy. Frequencies are calculated according to Eq.~(\ref{eq:f_0}), where $N_i$ is estimated by averaging of the corresponding stress on each resonator. The scale factor is calculated as a tangent to the linear dependence $\Delta f = f_1 - f_2$ on the applied acceleration $\hat{a}$ in the acceleration range $\pm 10$ (see Eq.~(\ref{eq:DeltafSF})); dependence of the SF on the voltage is shown in Fig.~\ref{fig:FEMDisplacements}(b). The scale factor for $V = 100$~(V) is estimated as 18.4~Hz/g that in the lumped plane capacitor model corresponds to the re-scaled voltage of $V = 100/\sqrt{156/3} = 14$~(V)  and  $SF = 20.7$~Hz/g. 

\section{Zero-voltage update and self-calibration}\label{sec:ZAU}
Since zero-signal updating algorithms show a significant improvement in the navigation systems' performance (see the Introduction for the examples), the simple built-in mechanism of detaching the sensing element from the input signal is a promising technique for the improvement of the inertial sensor figures of merit. In our case, if the voltages between the frames and the proof mass are zero,  there is no connection between the resonators and the inertial load. It works like an electrostatic bearing for the device and opens new possible scenarios for the reducing of the zero-bias due to the unpredictable noise, including the noise from signal processing electronic circuitry.

Furthermore, while detached from the inertial input, each of the resonators can be used as a sensor of some internal parameter of the device. For example, temperature: the vibrating beams together with the driving and signal processing electronic circuitry serve as a highly sensitive resonant temperature sensor. We show further that based on such measurements, the compensation of the sensitivity of the SF to temperature is possible. 

The dependence of the resonator's working frequency of  on the temperature was studied theoretically and in the FE model. 
We assume that the Young's modulus and linear dimensions of the device's component depend linearly on temperature $T$ \cite{Hopcroft2010WhatSilicon}:
\begin{equation}\label{eq:Ethermo}
    E(\theta) = (1-\gamma \theta) E_0, \quad \theta = T-\Tr,\quad \Tr = 293.15 \mathrm{K},\ \gamma = 79\ \mathrm{ppm/K}
\end{equation}
\begin{equation}\label{eq:L}
    L(\theta) = L_0(1+\alpha\theta),\quad \alpha = 2.6\ \mathrm{ppm/K}.
\end{equation}
Hereafter, $\theta=T-T_{ref}$ is an increase in the temperature above the ambient. We assume that the substrate holding the anchoring points of the suspension beams and of the resonators also undergoes the uniform thermal expansion, such that there are no internal thermal stresses in the absence of external loading.
The resonant frequency decreases linearly due to the changes of the Young's modulus and linear dimensions (note that we take into account the dependence of the density on the temperature, such that the mass is constant):
\begin{equation}
f_0(\theta)=\frac{\lambda_1^2}{2\pi}\sqrt{\frac{EI_t}{\rho A_tL^4}}=\frac{\lambda_1^2}{2\pi}\sqrt{\frac{E_0(1-\gamma \theta)I_{t0}(1+\alpha\theta)}{m L_0^3}} \approx f_0\left(1-\eta\theta\right),\ \eta = \frac{\gamma -\alpha}{2}
\label{eq:fotheta}
\end{equation}

\begin{figure}
 \centering
    \begin{minipage}[c]{.55\linewidth}
\includegraphics[width=0.95\textwidth]{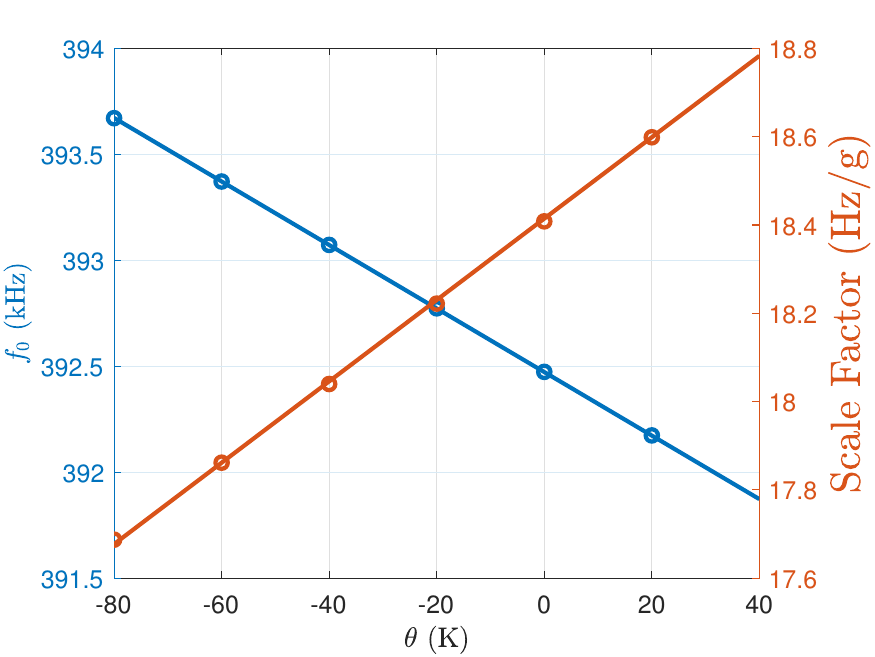}
    \captionof{figure}{Dependence of the resonator's working mode natural frequency (blue line) and of the scale factor (red line) on the temperature for the device with three electrodes. The tuning voltage is $V = 100\,(V)$.   The circles represent the FE model results, the lines depict their linear fit.
    }
    \label{fig:SF_vs_V_theta0}
    \end{minipage}\hfill
\begin{minipage}[c]{.45\linewidth} 
\captionof{table}{Reference scale factor and its temperature coefficient TCSF for different voltages, calculated based on FE model's results.}
\begin{tabular}{|p{1.8cm}|p{1.8cm}|p{1.8cm}|}
\hline
Voltage (V) & $SF_{ref}$ (Hz/g) & TCSF (ppm/K) \\
\hline
50          & 1.5               & 129          \\
75          & 4.2               & 161          \\
100         & 18.4              & 501         \\
\hline
\end{tabular}
\label{tab:TCSF}
\end{minipage}
\end{figure}

The analysis of the dependence of the resonator natural frequency on the temperature was performed on the FE model of one resonator and the corresponding frame that is equivalent to  zero voltage $V$ in the full model. The modal analysis with pre-stress was used, the temperature range was set $\theta\in[-80,40]$ (Fig.~\ref{fig:SF_vs_V_theta0}, blue line, left $y$-axis). The relative error between theoretical $\eta$ (Eq.~(\ref{eq:fotheta})) and the FE model's estimation of working frequency's temperature coefficient is less than $0.001$\%. This estimation is also in a good agreement with the  experimental results \cite{zhang2023thermal} where $\mathrm{TC}f_0 = 32$ ppm/K and $\gamma = 63.82$ ppm/K were reported.

 In the FE analysis of the full device, only the dependence of the Young's modulus on the temperature Eq.~(\ref{eq:Ethermo}) was taken into account; the thermal expansion was neglected. For the fixed voltage $V$, the scale factor linearly increases with  the temperature of the device $SF = SF_{ref}(1+ \mathrm{TCSF}\cdot \theta)$ (see Fig.~\ref{fig:SF_vs_V_theta0}, red line, right $y$-axis, $V = 100$ (V)). For other voltages, the results are listed in Tab.~\ref{tab:TCSF}. By the order of magnitude,  $\mathrm{TCSF}$ is in accordance with the theoretical estimation $\mathrm{TCSF} \approx 3\gamma/2 = 119$~(ppm/K) discussed in \ref{App:softening}. 
 
In view of the aforesaid, the following procedure for temperature compensation can be suggested. Prior to the device operation, the resonators calibration as a ``built-in'' thermometer is carried out in the known controlled temperature environment. 
Furthermore, the calibration of the scale factor' dependence on the voltage and temperature should be performed. 
During the device operation,  by switching off the voltage (or by varying the voltage in a controllable, pre-defined, manner), the frequency $f_0(\theta)$ is measured for the temperature $\theta$  extraction; by switching the voltage on, from the device differential output $\Delta f$ and ``thermal'' look-up table, we get the current acceleration $a$. 

In this scenario, both resonators work as built-in thermometers. Since the sensor incorporates two resonators located at a certain distance between them, the sensing beams could be used also for the temperature measurements in a situation with non-uniform distribution of the temperature, which is often the case.  This feature distinguishes the proposed method from the differential compensation of the temperature shift by common mode rejection. Furthermore, since the regions with large values of $a\theta$ are not reachable by the common mode reduction \cite{wang2023temperature}, the proposed direct measurement of temperature has the potential to increase the acceleration and temperatures working ranges.

\section{Summary and Discussion}
In the present work a new bias zeroing paradigm is introduced based on the ON/OFF physically switchable/tunable transmission of the mechanical signal from the input to the sensing element. Specifically, we present a novel architecture of the resonant accelerometer with the ES transmission.
The simplified lumped  and full scale thermo-electro-mechanical finite element multi-physics numerical model of the 
device is used to demonstrate the feasibility of the suggested approach and to estimate the device's performance.
The lumped model is formulated and examined for two approaches used to describe the electrostatic forces:  the parallel-plate capacitor approximation and the numerical model based of the BEM analysis, taking into account fringing ES fields. The former allows analytical qualitative investigation of the device's nonlinear behavior, including instabilities engendered by electrostatic nonlinearities and tuning of the scale factor by the voltage. The latter gives more accurate numerical estimations for the device performance and working range. 

In general, in the considered device, the frequency of the sensing resonator is influenced, apart from the acceleration, also by the tuning voltage $V$, temperature $\theta$, and other parameters (denoted by a vector $\mathbf{p}$), representing for example the contribution of the electronic circuit, packaging, environmental factors, such as vibrations and acoustics. It can be represented in the form
\begin{equation}
\label{eq:freqgeneric}
f(a,V,\theta,\mathbf{p}) = f_{\mathrm{REF}}(\theta,\mathbf{p})\sqrt{1+V^2\mathcal{F}_{ES}(V,\theta,\mathbf{p})+aV^2\mathcal{F}_{a}(a,V,\theta,\mathbf{p})}.
\end{equation}
Here 
$\mathcal{F}_{ES},\;\mathcal{F}_{a}$
are functions associated with ES force and the inertial force, respectively, and (see Eq.~(\ref{eq:fotheta}))
\begin{equation}
\label{eq:freqgenericref}
f_{\mathrm{REF}}(\theta,\mathbf{p}) = f_{0}(0)\mathcal{F}_{\theta}(\theta,\mathbf{p}),
\end{equation}
is the natural frequency of the resonator at zero voltage and zero acceleration measured by the same sensing circuit as the inertial signal. It is still affected by the temperature and the additional parameters $\mathbf{p}$.

An ability to tune the inertial signal opens new possibilities for the control of the sensor's measurement range and sensitivity.
Switching off the sensing element could be used as a built-in temperature sensor, allowing direct thermo-calibration and extending the working temperature range.
Since each resonator serves as a thermal sensor, our approach also covers the case of the non-uniform temperature distribution, distinguishing it from the temperature compensation by common mode rejection. Note also that while on-chip integrated temperature sensors are routinely used in inertial sensor packages as part of thermal compensation table lookup procedures, the key distinguishing feature of our design is that the temperature and the inertial input are measured using exactly the same resonant sensing element and the same  sensing electronics. As a result, the contribution of the circuit to the scale factor drifts are accounted for during the reference frequency measurement with the disconnected inertial signal.
Compared to the ZUPT technology, the instants of the signal zeroing here could be picked up independently of the carrying object's real movement. It opens the possibility for the advanced scheme of the signal processing for such architecture of the sensors.
We anticipate that the implementation of the suggested approach
enables 
new signal processing algorithms, which have the potential to significantly reduce bias instability-related errors.
Furthermore, our approach opens the option for precise non-differential measurement, which could potentially eliminate the frequencies' locking phenomena.
We believe also  that the suggested approach of the input signal zeroing is not limited to the inertial or resonant sensors and can be extended to a very large variety of sensors operated statically or dynamically.

\vspace{3mm}
{\bf Acknowledgement.} Dr. A.~Zobova acknowledges the Center for Integration in Science of the Israeli Ministry of Aliyah and Integration for partial funding of this research. S. Krylov is supported by the Henry and
Dinah Krongold Chair of Microelectronics, Israel.

 \appendix

\section{Lumped model analysis}
\label{App:lumpedanalytic}
One of the advantages of the simplest lumped representation of the device with the electrostatic force evaluated using the parallel capacitor model is that it can be to a large extent treated analytically.
In the framework of this model, the equilibrium equations of the sensors are (see Eqs.~(\ref{equi1cap_ud}),~(\ref{equi2cap_ud})):
\begin{eqnarray}\label{equicap1_A}
\hat{u}& =&\tilde{a}+\frac{\eta\beta}{1-\hat{v}_1} - \frac{\eta\beta}{1-\hat{v}_2}, \\
\label{equicap2_A}
\hat{v}_1 &=& \frac{\beta (1+\hat{u})}{(1-\hat{v}_1)^2},\\
\label{equicap3_A}
\hat{v}_2& = &\frac{\beta (1-\hat{u})}{(1-\hat{v}_2)^2}.
\end{eqnarray}
By solving Eq.~(\ref{equicap1_A}) for $\beta$ we obtain
\begin{equation}
\label{eq:betavsuandv}
\beta=\frac{\left(\hat{u}-\tilde{a}\right) \left(1-\hat{v}_1\right) \left(1-\hat{v}_2\right)}{\eta  \left(\hat{v}_1-\hat{v}_2\right)}.
\end{equation}
Substituting of Eq.~(\ref{eq:betavsuandv}) into Eqs.~(\ref{equicap2_A}),~(\ref{equicap3_A}) yields two coupled nonlinear algebraic equations in terms of $\hat{u},\;\hat{v}_1,\;\hat{v}_2$, which describe parametric (with $\beta$ serving as the parameter) curves (the equilibrium path at a specific acceleration $\tilde{a}$) in the $(\hat{u},\;\hat{v}_1,\;\hat{v}_2)$ space Fig.~\ref{fig:equilibrium_curves_cap1}(a).

Substitution of $\hat{u}$ from Eq.~(\ref{equicap1_A}) into Eqs.~(\ref{equicap2_A}) and~(\ref{equicap3_A}) yields
\begin{eqnarray}\label{eqv1v2full1A}
\hat{v}_1 &=& \frac{\beta}{(1-\hat{v}_1)^2}\left(1+\tilde{a}+\frac{\eta\beta}{1-\hat{v_1}} - \frac{\eta\beta}{1-\hat{v_2}}\right),\\
\label{eqv1v2full2A}
\hat{v}_2& = &\frac{\beta}{(1-\hat{v}_2)^2}\left(1-\tilde{a}-\frac{\eta\beta}{1-\hat{v_1}} + \frac{\eta\beta}{1-\hat{v_2}}\right).
\end{eqnarray}
Eqs.~(\ref{eqv1v2full1A}),~(\ref{eqv1v2full2A}) describe the bifurcations diagram $\beta=\beta(\hat{v}_1,\hat{v}_2)$ (the equilibrium curves), Fig.~\ref{fig:equilibrium_curves_cap2}(a). Note that in general the formulation 
Eqs.~~(\ref{equicap1_A}),~(\ref{equicap2_A}),~(\ref{equicap3_A}) contains three variables $\hat{u},\;\hat{v}_1,\;\hat{v}_2$ and two parameters $\tilde{a}$ and $\beta$ and Figs.~\ref{fig:equilibrium_curves_cap2},~~\ref{fig:equilibrium_curves_cap1} depict certain cross-sections corresponding to the cases when two out of five parameters have fixed value.

We analyse now the frequency shift and the SF of the device.
At zero acceleration and in the case of a full symmetry  $\hat{u}=0$ and $\hat{v}_1=\hat{v}_2=\hat{v}_*=\beta(1-\hat{v}_*)^{-2}$. Set in Eqs.~(\ref{eqv1v2full1A}),~(\ref{eqv1v2full2A}), $\hat{v}_1=\hat{v}_*+\hat{w}_1,\;\hat{v}_2=\hat{v}_*-\hat{w}_2$, where $\hat{w}_1,\;\hat{w}_2$ are displacements of the frames due to the acceleration and replace $\beta=\hat{v}_*(1-\hat{v}_*)^2$. Linearizing  of the outcome for $\hat{w}_i \ll 1$, we obtain
\begin{eqnarray}
\label{eqw1linA}
\frac{1+\eta\hat{v}_*^{3} -\eta \hat{v}_*^{2}  -\left(2 \tilde{a}+3\right) \hat{v}_*}{1-\hat{v}_*}\hat{w}_1
-\eta\hat{v}_*^{2}  \hat{w}_2=\tilde{a} \hat{v}_*,\\
-\eta\hat{v}_*^{2}  \hat{w}_1+\frac{1+\eta\hat{v}_*^{3}  -\eta\hat{v}_*^{2}  +\left(2 \tilde{a}-3\right) \hat{v}_*}{1-\hat{v}_*}\hat{w}_2=\tilde{a} \hat{v}_*.
\label{eqw2linA}
\end{eqnarray}
The system of two linear (in terms of $\hat{w}_1,\;\hat{w}_2$) equations~(\ref{eqw1linA}),~(\ref{eqw2linA}) can be solved resulting in the  expressions for the displacements $\hat{w}_i=\hat{w}_i(\tilde{a},\hat{v}_*(\beta)),\;\;i=1,2$ which are nonlinear functions of the accelerations and the voltage.

Consider now the relative frequency shift  (see Eq.~(\ref{eq:f_0}))
\begin{equation}
\frac{\Delta f}{f_0} = \sqrt{1+N_1/N_E}-\sqrt{1+N_2/N_E}
\label{eqfreqshiftA}
\end{equation}
In view of Eq.~(\ref{eq:axialforce}) and recalling that $k_t=EA_t/L,\;N_E=4\pi^2EI_t/L^2$ we obtain
\begin{eqnarray}
\label{eq:N1vswnondim}
\frac{N_1}{N_E} &=& \frac{k_t({v}_*+w_1)}{\mathcal{A}_0N_E} = \frac{\varepsilon\hat{v}_*}{4\pi^2 \hat{r}^2\mathcal{A}_0}+\frac{\varepsilon \hat{w}_1}{4\pi^2 \hat{r}^2\mathcal{A}_0},\\
\label{eq:N2vswnondim}
\frac{N_2}{N_E} &=& \frac{k_t({v}_*-w_2)}{\mathcal{A}_0N_E} = \frac{\varepsilon\hat{v}_*}{4\pi^2 \hat{r}^2\mathcal{A}_0}-\frac{\varepsilon \hat{w}_2}{4\pi^2 \hat{r}^2\mathcal{A}_0},
\end{eqnarray}
where $\varepsilon=g_0/L,\; \hat{r}=\sqrt{I_t/(A_tL^2)}$. 
Equation~(\ref{eqfreqshiftA}) with $N_1/N_E,\,N_2/N_E$ expressed in terms of the frame displacements $\hat{w}_1,\;\hat{w}_2$ relates to the frequency shift, the acceleration (through the displacements $\hat{w}_1,\;\hat{w}_2$) and the zero acceleration displacement $\hat{v}_*=\hat{v}_*(\beta)$ due to the electrostatic force.  Our calculations show that consistently with Eq.~(\ref{eq:dfvsalin}) the SF grows nonlinearly with $\hat{v}_*$ (and therefore with voltage parameter $\beta$, see Fig.~\ref{fig:SF_cap}(a)). 

To estimate the dependence of the frequency shift and therefore of the SF on the voltage, we substitute Eqs.~(\ref{eq:N1vswnondim}),~(\ref{eq:N2vswnondim}) into Eq.~(\ref{eqfreqshiftA})
and linearize the outcome for $\tilde{a} \ll 1$
\begin{equation}
\label{eq:dfvsalin}
\frac{\Delta f}{f_0} \approx  \frac{ \varepsilon \tilde{a} \hat{v}_* \left(1-\hat{v}_*\right)}{2\pi\hat{r}\sqrt{\mathcal{A}_0}(1-3 \hat{v}_*-2 \hat{v}_*^{2} \eta+2\hat{v}_*^{3} \eta)\sqrt{\varepsilon \hat{v}_*+4\pi^2\hat{r}^2\mathcal{A}_0}}, \qquad SF_{0} = \frac{d}{d\tilde{a}}\left(\frac{\Delta f}{f_0}\right).
\end{equation}
When $\hat{v}_*$ approaches the critical value $\hat{v}_{*}^{crit}$, the  scale factor $SF$ grows to infinity (note that $\hat{v}_{*}^{crit}$ can be found from the condition that the denominator of Eq.~(\ref{eq:dfvsalin}) is zero).
For small voltages $\hat{v}_* \ll 1$, we have $\hat{v}_*=\beta(1-\hat{v}_*)^2 \approx \beta$, and Eq.~(\ref{eq:dfvsalin}) can be simplified
\begin{equation}
\label{eq:dfvsasmallvoltage}
\frac{\Delta f}{f_0} \approx  \frac{ \varepsilon \tilde{a} \hat{v}_* }{4\pi^2\hat{r}^2\mathcal{A}_0} = \frac{ \varepsilon \tilde{a} \beta}{4\pi^2\hat{r}^2\mathcal{A}_0}.
\end{equation}
Since $\beta = \dfrac{n\epsilon_0 b  d_{0e}}{2 \keff g_0^3 }\,V^2$ and $k_{\mathrm{eff}}
= \dfrac{k_t}{\mathcal{A}_0^2} \left(1+\dfrac{\mathcal{A}_0^2\kfs}{k_t}\right)$ (see Eqs.~(\ref{eq:keff}),~(\ref{eq: dimensionless_pars})), so $\beta \propto \mathcal{A}_0^2 V^2$. One may conclude therefore that at small voltage limit the SF is proportional to the product of the geometric amplification and square of the tuning voltage. 
\section{The influence of the material thermo-softening on the scale factor}
\label{App:softening}
Here we study how the change of the Young's modulus with temperature (Eq.~\ref{eq:Ethermo}) influences the scale factor of the device. To simplify matters and highlight one of the calibration scenarios of the device, we neglect the thermal stresses in the resonators and the suspensions beams, along with the change of the gaps between the electrodes due to thermal expansion. In the framework of the lumped model's level, the dependence of the stiffness of the structural elements on the temperature is solely due to the Young's modulus thermal dependence, such that $k(\theta) \approx \xi k_0$ (where $k$ is either $k_t,\;k_h,\;k_f$) for $\xi = 1-\gamma\theta$ and $\gamma$ defined in Eq.~(\ref{eq:Ethermo}). 

Note that for any dependence of the electrodes' capacitance on the coordinates $u, v_i$, the equations of equilibrium Eqs. (\ref{equi1}),~(\ref{equi2}) are invariant under simultaneous scaling of all the stiffnesses $k$, the input acceleration and square voltage
\begin{equation}
k \longrightarrow \xi k,\ 
a \longrightarrow \xi a,\ V^2 \longrightarrow \xi V^2
\end{equation}
for any  $\xi$. That means that if we denote 
the dependence of the equilibrium states on the input acceleration $a$, squared voltage and temperature increase $\theta$ as $u(a,V^2,\theta)$ and $v_i(a,V^2,\theta)$, this invariance of equations assures that
\begin{equation}
v_i( a, V^2,\theta) = v_i\left(\frac{a}{\xi},\frac{V^2}{\xi},0\right),\quad
u(a,V^2,\theta) = u\left(\frac{a}{\xi},\frac{V^2}{\xi},0\right).
\end{equation}
The acceleration-induced axial forces acting on the beam, Eq.~(\ref{eq:axialforce}), are proportional to the stiffness of the resonator $k_t$ and the frames' displacements $v_i$, hence 
\begin{equation}
    N_i(a,V^2,\theta) = \xi N_i\left(\frac{a}{\xi},\frac{V^2}{\xi},0\right).
\end{equation}
When the voltage and the acceleration are both non-zero, the frequency is 
\begin{equation}
f_i(a,V,\theta) = f_{0}(\theta)\sqrt{1+ \frac{N_i(a,V,\theta)}{N_E(\theta)}},
\end{equation}
where $f_0(\theta)$ and $N_E(\theta)$ are the temperature-dependent natural frequency (see Eq.~(\ref{eq:fotheta})) and the buckling force of the double-clamped beam. After the linearization, we get the frequency shift as 
\begin{eqnarray}
&\Delta f(a,V^2,\theta) =f_0(\theta)\dfrac{N_1(a,V,\theta)-N_2(a,V,\theta)}{2N_E(\theta)} \\
&=  f_{0}\left(1-\eta\theta\right)\dfrac{\xi N_{1}\left(\dfrac{a}{\xi},\dfrac{V^2}{\xi},0\right)-\xi N_{2}\left(\dfrac{a}{\xi},\dfrac{V^2}{\xi},0\right)}{2\xi N_E(0)}  = {(1-\eta\theta)}\, \Delta f_{\mathrm{ref}}\left(\dfrac{a}{\xi},\dfrac{V^2}{\xi},0\right),&
\label{eq:DeltafT}
\end{eqnarray}
where $\Delta f_{\mathrm{ref}}$ is the dependence of the frequency shift on acceleration and voltage at the reference temperature.

Assume that for all fixed $\theta$ within the working range the SF nonlinearity is neglectable, then, according to Eq.~(\ref{eq:DeltafSF}), $\Delta f = \hat{a}\,SF$, and therefore 
\begin{equation}
SF(V^2,\theta) = \frac{1-\eta\theta}{\xi}{SF_{\mathrm{ref}}\left(\frac{V^2}{\xi}\right)}
\end{equation}
Taking into account that the scale factor for small voltages is proportional to $V^2$ (see Eq.~(\ref{eq:dfvsasmallvoltage})), 
we get
\begin{eqnarray}
\label{eq:TCSF}
SF(V^2,\theta)&=& 
\frac{1-\eta\theta}{\xi^2} SF_{\mathrm{ref}}(V^2) = 
(1+(2\gamma-\eta)\theta) SF_{\mathrm{ref}}(V^2) 
\\
&=&\left(1+\frac32\gamma\theta\right) SF_{\mathrm{ref}}(V^2) 
\end{eqnarray}
(the last equality is based on the estimate  of the sensing frequency temperature coefficient $\mathrm{TC}f_0 = \eta \approx \gamma/2$, Eq.~(\ref{eq:fotheta})).

This formula explains the increasing of the SF with the temperature and gives a rough estimation for its temperature coefficient $\mathrm{TCSF}$. By the derivation procedure, it holds within the range of acceleration, voltage and temperature defined by negligible nonlinearity of the scale factor, and linear dependence of the SF on the squared voltage.
Furthermore, we note that the $\mathrm{TCSF}$ is impacted also by the effects, omitted here: such as the narrowing of the gaps between the electrodes due to the thermal expansion of the material and thermal stresses in the resonators and the suspension beams.


\end{document}